\documentclass[12 pt]{article}
\usepackage{amsmath,amssymb}
\usepackage{graphicx, color, array}
\usepackage[margin = 1in]{geometry}

\usepackage{subfigure}

\newtheorem{Definition}{Definition}[section]
\newtheorem{Theorem}{Theorem}[subsection]
\newtheorem{Lemma}{Lemma}[section]
\newtheorem{Proposition}{Proposition}[section]

\newcommand{\transp}{^{\mbox{\scriptsize \sf T}}}

\usepackage{chngcntr}

\begin{document}
\title{Evolving Network Model that Almost Regenerates Epileptic Data}
\author{G.~Manjunath  \\
{\small \it Department of Mathematics, Rhodes University, South Africa} \\ {\small Email: $\{\mbox{m.gandhi}\}$@ru.ac.za } \\}

\date{}
\maketitle

\begin{abstract}
In many realistic networks, the edges representing the interactions between the nodes are time-varying.  There is growing evidence that the complex network that models the dynamics of the human brain has time-varying interconnections, i.e., the network is evolving. Based on this evidence, we construct a patient and data specific evolving network model (comprising discrete-time dynamical systems) in which epileptic seizures or their terminations in the brain are also determined by the nature of the time-varying interconnections between the nodes. A novel and unique feature of our methodology is that the evolving network model remembers the data from which it was conceived from, in the sense that it
evolves to almost regenerate the patient data even upon presenting an arbitrary initial condition to it. We illustrate a potential utility of our methodology by constructing an evolving network from clinical data that aids in identifying an approximate seizure focus -- nodes in such a theoretically determined seizure focus are outgoing hubs that apparently act as spreaders of seizures.  We also point out the efficacy of removal of such spreaders in limiting seizures.
\end{abstract}


\section{Introduction}
Increasingly, many complex systems are being modeled as networks since the 
framework of nodes representing the basic elements of the system and the interconnections of the network representing the interaction between the elements fits well for a theoretical study. When the complex systems are large-dimensional dynamical systems, the network framework comprises many interacting subsystems of smaller dimension each of which constitutes a node. As a particular example, the whole or a portion of the entire neuronal activity in the human brain can be regarded as the consequential dynamics of interacting subsystems, where the dynamics of a subsystem is generated by a group of neurons. The enormously interconnected subsystems in the brain generate a wide variety of dynamical patterns, synchronised activities and rhythms.

Epilepsy is a disorder that affects the nerve cell activity which in turn intermittently causes seizures. During such seizures, patients could experience  abnormal sensations including loss of consciousness. Clinical and theoretical research have shown that underpinning the cause of this disorder has not proved to be easy, in particular the predicament of resolving the question as to whether the  disorder manifests due to the nature of the interconnections in the brain or the pathologicity of a portion of the brain tissue itself remains.  

Since epilepsy is one of the most common neurological disorders with an estimate of more than 50 million individuals being affected \cite{Ngugi}, there is a strong need both for curative treatment and as well for the analysis of the underlying structure and dynamics that could bring about seizures. In this paper, our attention is on focal epilepsy where the origin of the seizures are circumscribed to certain regions of the brain called the seizure focus \cite{bragin2000, bonilha2012,diessen2013,kramer2012,richardson2012,spencer2002,terry2012}, and the aim of this paper is to foster theoretical investigation into the connection strengths of the underlying nodes in such a seizure focus in comparison to the other nodes in the network.

Different models have been proposed to understand different aspects of focal epilepsy \cite{benjamin2012,jansen1995,wendling2005}.  Mathematically speaking, dynamical system models of focal epilepsy studied in the literature are mainly of two broad categories: (i). Models that consider noise to be inducing seizures in a node \cite{kalitzin2010,silva2003,  suffczynski2004} (ii). Models that consider seizures to be caused or terminated by a bifurcation, a change in the intrinsic feature of the (autonomous) dynamics at the node \cite{breakspear2006, marten2009,robinson2002}.

While modeling the neuronal activity of the human brain, there are far too many numerous parameters which also dynamically evolve on separate spaces and scales, and it is unrealistic to observe or encapsulate all these aspects in a static network or an autonomous dynamical system (see Appendix~\ref{AppendixAA}).  Since the brain evolution does not depend on its own internal states as it responds to external stimuli and the interconnections in the brain are likely to change in response to stimuli, the human brain can be safely said to be a nonautonomous system and in the language of network dynamics, this is effectively an evolving network (see Section~\ref{sec_SFN}).

Besides evidence that network connections play a role in initiating seizures (e.g.,
\cite{bertram1998,bragin2000,Lemieux2011,spencer2002}), some authors also provide evidence that stronger interconnections play a role in epileptic seizures (e.g., \cite{HutchingsHKWTK15,morgan2008nonrandom}). Also, a commonly found feature of biological systems is adaptation: an increase in exogenous disturbance beyond a threshold  activates a change in the physiological or biochemical state, succeeded by an adaptation of the system that facilitates the gradual relaxation of these states toward a basal, pre-disturbance level. 
Based on all these, we present a novel phenomenological model where seizures could be activated in a node, if certain interconnections change to have larger weights resulting in a larger exogenous drive or disturbance or energy transmitted from other nodes into it, and upon experiencing seizure,  adjustment of interconnections takes place eventually to subside the seizures. Since the interconnections are allowed to change, the model dynamics emerges from an evolving network.

In this paper, we propose a patient and data-specific, macroscopic, functional network model that can
act as a ``surrogate" representation of the electroencephalography (EEG) or electrocorticography (ECoG) time-series.
The network model is new in the sense that it satisfies the following simultaneously:  the network connections are directed and weighted; the network connections are  time-varying; there is a discrete-time dynamical system at each node whose dynamics is constantly influenced by the rest of the network's structure and dynamics; the ability of the network to generate a time-series that sharply resembles the original data is based on a notion of robustness of the dynamics of the network to its initial conditions what we call as the ``state-forgetting" property. 

The essence of the state-forgetting property, roughly put is that the current state of the brain does not have an `` distinct imprint" of the brain's past right since  from its ``genesis" or ``origin" (rigorous definitions are provided later in the paper). Effectively, this means that the brain tends to forget its past history as time progresses. Such a notion has been adapted from the previously proposed notions for studying static (artificial neural) networks with an exogenous input in the field of machine learning viz., echo state property  \cite{jaeger2001, manjunath2013}, liquid state machines, \cite{maass2002}, backpropagation-decorrelation learning,\cite{steil2004} and others. However, the network model that we represent is very different from that of the artificial neural networks and moreover the nodes in our model correspond to different regions of the brain.

Our methodology is as follows: We present an evolving network model which exhibits state-forgetting property for all possible inter-connections, and a particular sequence of  interconnection strengths (matrices) is determined from a given EEG/ECoG time-series so that the model can almost regenerate the time-series.  
The weights of the interconnections are chosen from a set of possibilities based on a heuristic algorithm that infers a causal relationship between nodes  -- a (directed) synchrony measure between two nodes $i$ and $j$ represents the conductance (ease at which energy flows) from node $j$ into node $i$, and  weight of the connection from node $j$ into node $i$ has some degree of proportionality with the synchrony measure.

The state forgetting property ensures the synthesized evolving network has in some sense captured the original time series as an ``attractor" (see \cite{manjunath2014} and Proposition~\ref{prop_pullback}), and hence the network acts as surrogate representation of the original time series. In other words, since any initial condition presented to a state-forgetting network has no effect on its asymptotic dynamics, and that the asymptotic dynamics resembles the time-series, we assume that the evolving network model has captured the essence of the time-series data. 

To indicate the potential utilities of the model, we analyse the constructed evolving network from a patient's clinical data. 
By comparing the relative strength of the interconnections, we identify nodes that are outgoing hubs of the network as a theoretically determined seizure focus. The
nodes in such a focus have a strong location-wise correlation with the set of nodes in a clinically determined  seizure focus. We also, discuss the efficacy of surgical removal of the determined focus in subsiding seizures.

The remainder of the paper is organised as follows. We introduce the notion of state-forgetting evolving networks in Section~\ref{sec_SFN}; formal definitions are made in Appendix~\ref{AppendixA}.  
In Section~\ref{sec_Model}, we present the evolving network model in detail and state a result concerning the state-forgetting property; the complete mathematical proof of this result is presented in Appendix~\ref{AppendixB}. In Section~\ref{sec_Weights}, we present a method to obtain the weight matrices to obtain a time-varying network from a clinical time series. In Section~\ref{Sec_results}, we present some results on the network inference followed by a brief discussion and conclusion in Section~\ref{sec_conclusion}.

\section{Evolving networks that are state-forgetting} \label{sec_SFN}

The aim of this section is to give an intuitive description of the state forgetting property of evolving networks; the formal and rigorous definitions of this property are found in Appendix~\ref{AppendixA}. We employ discrete-time systems \cite{devaney1989} since they are more amenable to computer simulations as the computed orbits or solutions unlike in the case of differential equations do not depend on numerical methods for computing the solutions and their related accuracy.  Also, discrete-time systems can exhibit complex dynamical behaviour (e.g., \cite{devaney1989}) even in one-dimension whereas a continuous time-system needs a dimension of at least three. 

\begin{figure}[!ht]
\centering
\includegraphics[scale=0.37]{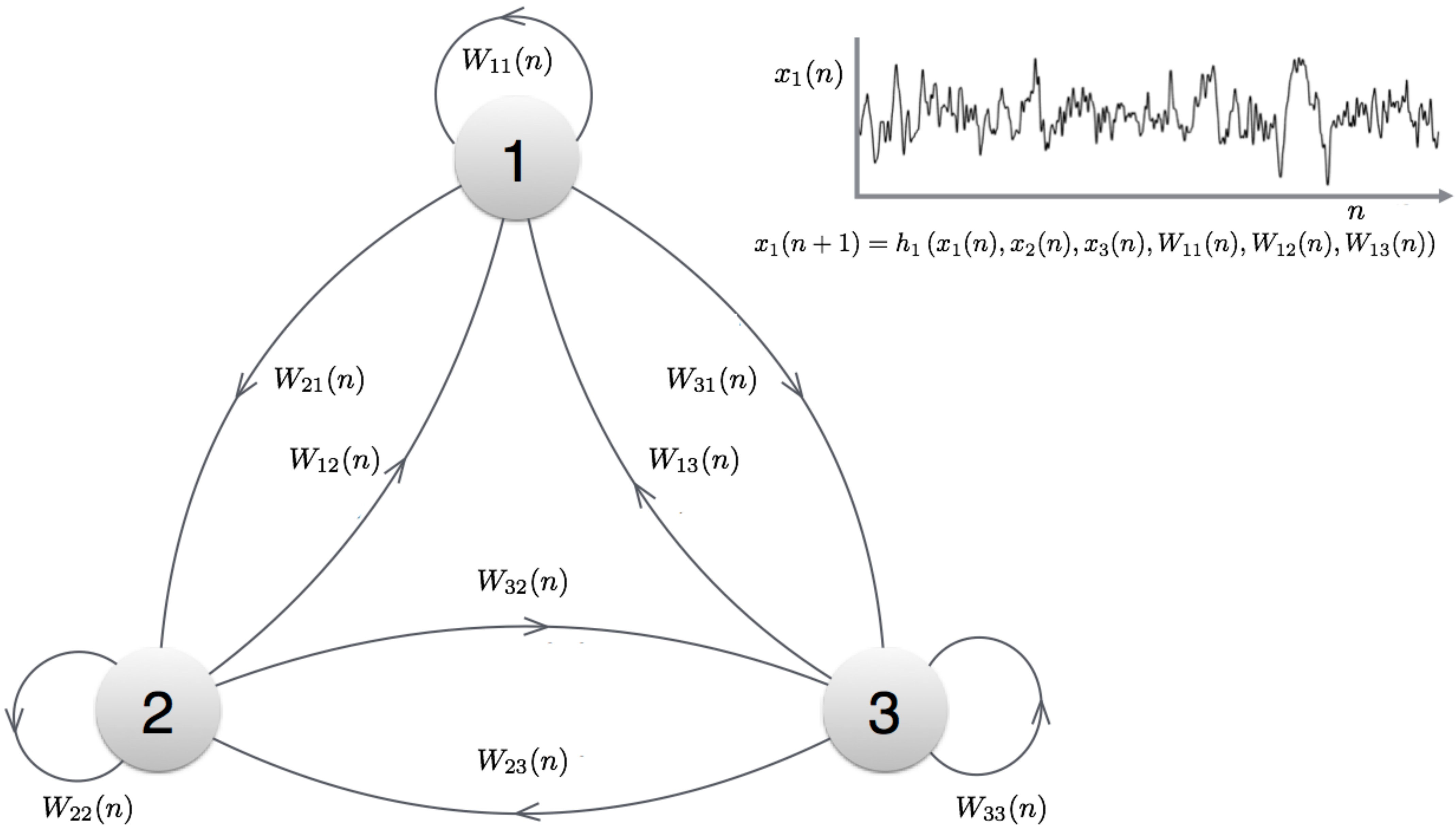}
\caption{A schematic of an evolving network with three nodes and the plot of the time series $x_1(n)$ from Node $1$} \label{fig_3nodes}
\end{figure}

Using Fig.~\ref{fig_3nodes}, we
present a less formal and simplistic description of an evolving network
with three nodes $i=1,2,3$.  Let $W(n)$ be a $3 \times 3$ matrix defined for each integer $n$ with $W_{ij}(n)$ being the strength of the incoming connection from node $j$ to node $i$; here $W_{ij}(n)$ is the element in the $i^{\mbox{th}}$ row and $j^{\mbox{th}}$ column of $W(n)$. 
There is a discrete-time dynamical system at each node in the evolving network represented in Fig.~\ref{fig_3nodes}, and the $i^{\mbox{th}}$ node dynamics is given by
$$
x_i(n+1) = h_i\big(x_1(n),x_2(n),x_3(n),W_{i1}(n), W_{i2}(n),W_{i3}(n)\big),
$$ 
where $h_i$ is some function (not dependent on $n$) that calculates $x_i(n+1)$ using information from the dynamical state in the three nodes at time $n$ and the respective incoming  strengths into node $i$.
By specifying $h_1$, $h_2$ and $h_3$ together with the sequence of weight matrices $W(n)$, the evolving network is specified completely. By denoting $x(n) := (x_1(n), x_2(n), x_3(n))$, we can represent the evolving network dynamics also as a nonautonomous system $x(n+1) = g_n(x(n))$, where the variation of $g_n$ with $n$ represents the inherent plasticity of the system (change in $W(n)$ above). We can also allow $g_n$'s to represent external stimuli to the network and other obfuscated factors.

Since an evolving network is essentially a nonautonomous system, we recall some relevant results of a nonautonomous system. 
Of interest to us is the notion of an entire-solution of a nonautonomous system. An entire solution (or orbit) of a nonautonomous system  $\{g_n\}$ is a sequence $\{\vartheta_n\}$ that satisfies $\vartheta_{n+1} = g_n(\vartheta_n)$ for all $n \in \mathbb{Z}$.  It is possible to show under natural conditions that there exist at least one entire solution \cite{manjunath2013}. In the particular case, where the evolving network has a unique entire solution, we say that the network (or the nonautonomous system) has the state-forgetting property (formally defined in Definition~\ref{def_SF}).

Now, we comment on the rather intriguing
consequence of an evolving network $\{g_n\}$ having multiple entire-solutions. Suppose $\{\vartheta_n\}$ and $\{\theta_n\}$ are two distinct entire-solutions of a nonautonomous dynamical system, then for some time $n=n_0$, we have $\vartheta_{n_{0}} \not= \theta_{n_{0}}$ i.e, the system could have evolved into two possible states at time $n_0$.  Since $\vartheta_{n_{0}} = g_{n_{0}-1}(\vartheta_{n_{0}-1})$ and $\theta_{n_{0}} = g_{n_{0}-1}(\theta_{n_{0}-1})$, it is necessary to have $\vartheta_{n_{0}-1} \not= \theta_{n_{0}-1}$. Similarly, $\vartheta_{n_{0}-1} \not= \theta_{n_{0}-1}$ would necessitate $\vartheta_{n_{0}-2} \not= \theta_{n_{0}-2}$ and so on. Thus, in general, at any time-instant $n_0$, $\vartheta_{n_{0}} \not= \theta_{n_{0}}$ implies that there are two entire-solutions that have had distinct infinitely long pasts. In other words, the dynamics of such systems with distinct entire-solutions ``depends" not only on the inherent plasticity or the exogenous stimuli represented by the maps $\{g_n\}$, the distinct solutions carry a distinct and non-vanishing imprint of their past.

From the arguments in the preceding paragraph, for nonautonomous systems like the human brain, having more than one entire solution translates to saying that the current state of the brain is not just the result of the changing interconnections in the brain or the 
exogenous input but to also say that its past states have a non-vanishing effect on the asymptotic dynamics.

Since it is hard to envisage or believe that seizures in current time had something to do with the state of the brain right from its ``genesis" (genesis is used as a metaphor to refer to the earliest factual time instant from which the current brain state has evolved from, and this time could be matter of debate), we assume the more plausible -- seizures at a certain time are a consequence of how the network structure has evolved/changed till that time or a result of a stream of exogenous inputs to the brain, but certainly not a consequence of the state at which the brain was in during its genesis. This also forms the basis for employing echo-state networks in artificial neural network training and machine learning\cite{jaeger2001,manjunath2013}.

Whenever a nonautonomous system (technically, defined on a compact state space) has the state-forgetting property, it also has    state-space contraction towards its entire-solution in the backward direction (see Proposition~\ref{prop_pullback}). In fact, a stronger result \cite{manjunath2014} holds: if $\{x_n\}$ is the unique entire solution, then for every $y_0 \in X$ the successive iterates $y_{n+1} = g_n(x_n)$ would satisfy $|y_n-x_n| \to 0$ as $n \to \infty$. In other words, $\{x_n\}$ ``attracts" all other state-space iterates towards it (component-wise) as time tends to infinity. This convergence due to the attractor property of the state-forgetting property is what makes our evolving network model generate a time-series that actually sharply resembles (mathematically, converges to) the data from which the network was conceived.  Further discussion on attractivity is beyond the scope of this paper. 
The larger point is that for an evolving network to be of utility as a surrogate representation of the time-series data, we need the state-forgetting property.

\section{Model} \label{sec_Model}

Before, we actually describe the evolving network model in \eqref{eq_FullModel_New}, we list and describe its features: {\bf (i). }
The evolving network model has $N$-nodes for all time. {\bf (ii). } Placed at each node is a certain bi-parametric one-dimensional discrete-time dynamical system which is autonomous when there are no incoming connections  from other nodes to it, and nonautonomous otherwise. We refer to the dynamics of the system at a node, as  the dynamics of the node or nodal dynamics. {\bf (iii). } Among the two parameters that specify this one-dimensional dynamical system is a stability parameter. The nodal dynamics in autonomous mode has a globally stable fixed point at $0$ whenever the stability parameter is in a certain range (stable-regime); a globally stable fixed point attracts all solutions, i.e, the time series values measured in the node are close to $0$ and tend towards to it. 
{\bf (iv). } When the stability parameter is chosen 
not to lie in the stable regime, the autonomous  nodal dynamics can render  a range of different behaviours including  oscillatory behavior, complex behavior like chaos \cite{devaney1989} depending upon the value of the stability parameter. {\bf (v). } When the nodal dynamics is nonautonomous, the dynamics of the node is perturbed at each time-instant from the dynamical states of other nodes. We call the net perturbation from other nodes as the drive to the node under consideration. The exogenous drive from other nodes has its signature on the nodal dynamics, and if sufficiently strong, is also capable of setting the nodal dynamics  far away from the fixed point $0$. {\bf (vi). } We term a seizure in the node to be a dynamical excursion away from the fixed point, and that the seizures in a node are brought about (and terminated) in two distinct scenarios: (via). changing the stability parameter of the node in the autonomous mode of operation (vib). by the nature of the exogenous drive from the other nodes in the network in the nonautonomous mode. 
{\bf (vii). } In this paper, we consider the case of nodal dynamics slipping into seizures as in scenario (vib) above, i.e, we consider each node to have its stability parameter to lie in the stable regime, and the exogenous drive to cause seizures. The drive is time-varying on two accounts: first, being a function of the states of the other nodes it is time-varying as other nodes have a dynamical system and not static; second, the drive is time-varying since it is a function of the interconnection strengths of the nodes.

Our model is also motivated on the lines of previous research \cite{jirsa2014, schmidt2014,taylor2013, wendling2002} showing that the hyper-interconnected regions in the brain are prone to seizures (see Section~\ref{sec_Model}). Hence, we expect the drive to an individual node is likely to be having large magnitude if it has large weighted incoming connections.

\begin{center}
\begin{figure}[h]
\centering
\subfigure[$f_{a,b}$ with $a=0.01$ and $b=0.9$]{\includegraphics[scale=0.35]{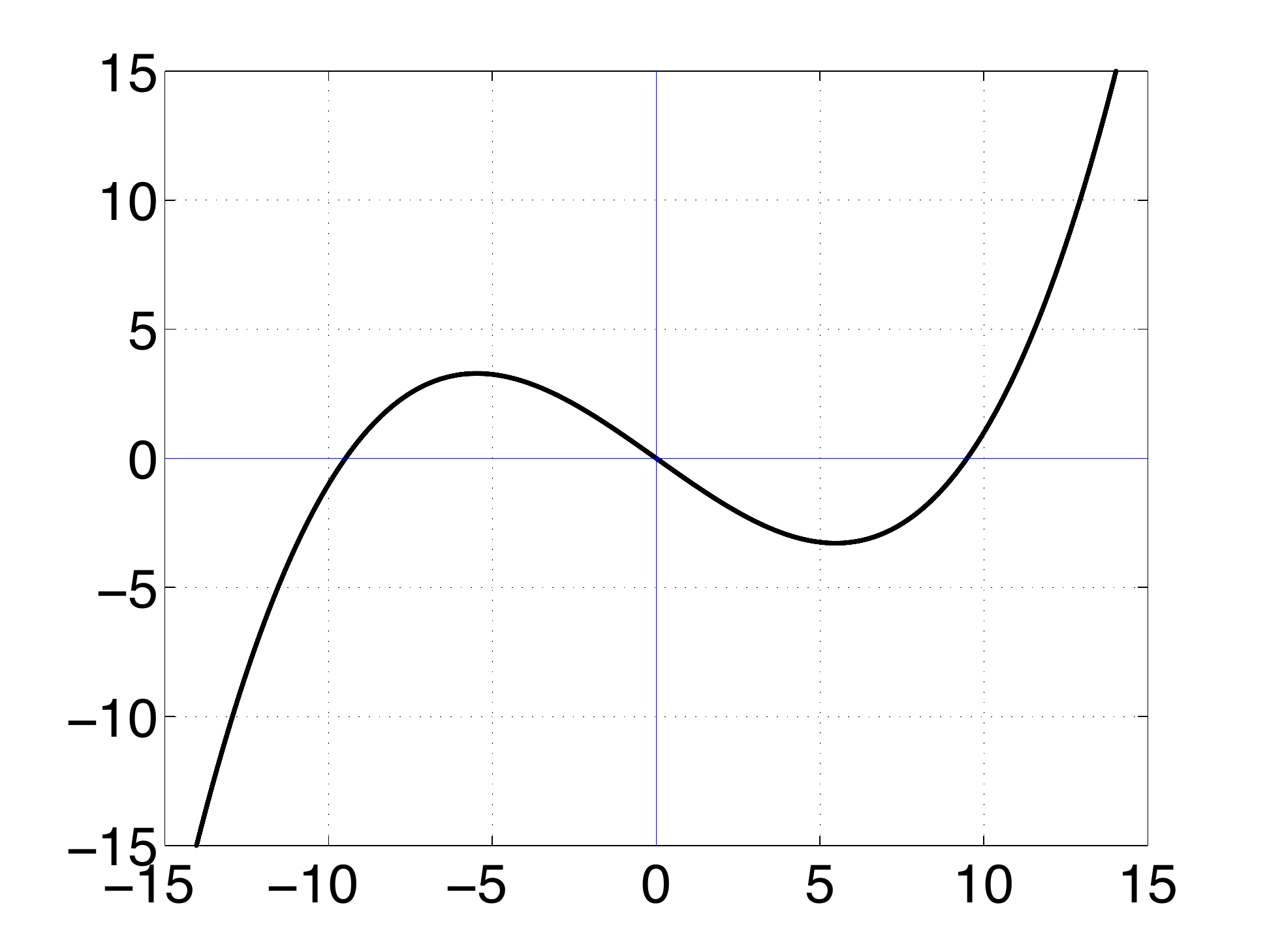}\label{figa}}
\centering
\subfigure[$\sigma$ with $p=5$ and $r=5$]{\includegraphics[scale=0.35]{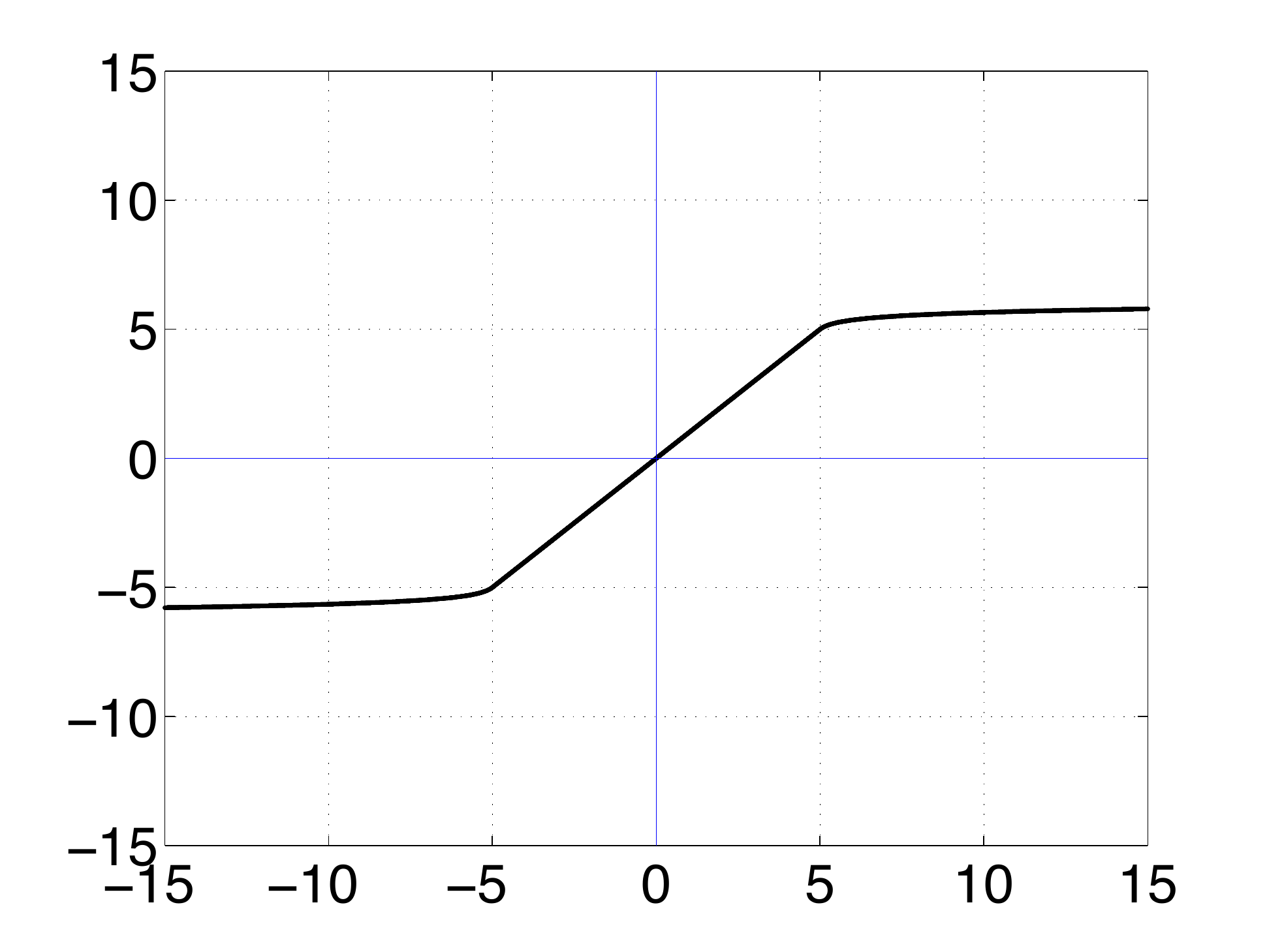}\label{figb}}
\centering
\subfigure[$\phi$ with $d=500$]{\includegraphics[scale=0.35]{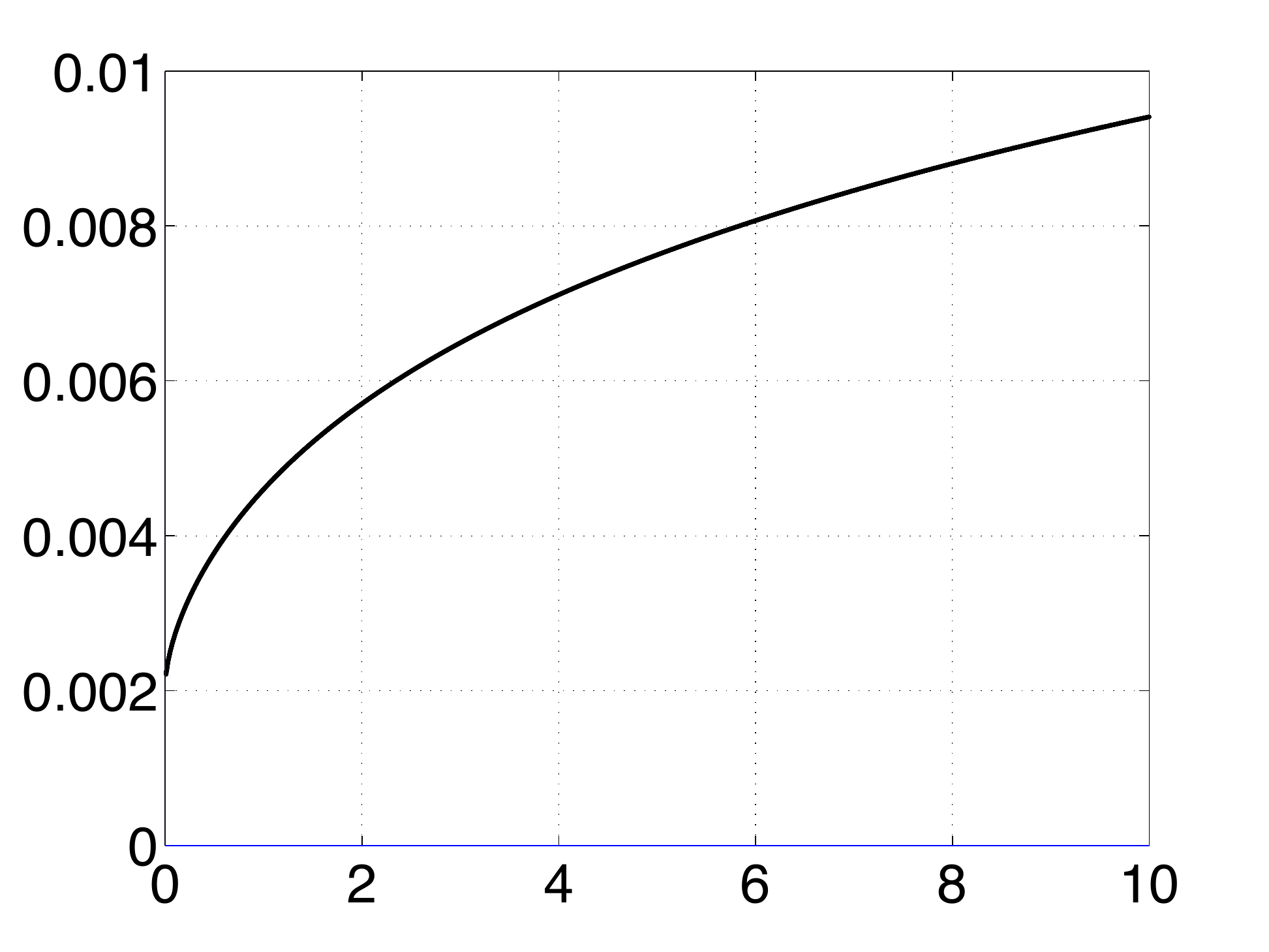}\label{figc}}
\caption{Graphs of $f_{a,b}$, $\sigma$ and $\phi$}
\end{figure}
\end{center}

To describe a family of evolving networks (as in Definition~\ref{def_EN}), we first sketchily
describe the nonautonomous dynamics at each node $i$ (the map $h_i$ described in Section~\ref{sec_SFN} or in Definition~\ref{def_EN}) via the update equation:
\begin{align} \label{eq_sketchy}
 \left \lgroup \mkern-5mu \begin{array}{ccc} x_i(n+1), \mbox{ state} \\
 \mbox{of node at} \\ \mbox{time } n+1 \end{array} \mkern-5mu \right \rgroup = 
& 
\left. \begin{array}{cccc} \: \\ \mbox{saturation} \\ \mbox{function} \\ \: \end{array} \mkern-10mu \right (  \mkern-5mu
\left \lgroup \begin{array}{ccc} f_{a,b} (x_i(n)), \mbox{the} \\
 \mbox{update at the } \\ i\mbox{th node}  \\
 \end{array} \right \rgroup + 
\left. \left \lgroup \begin{array}{ccc} \mkern-5mu \mbox{net drive from} \\ \mbox{from all other} \\
 \mbox{nodes } j\not=i 
 \end{array} \mkern-5mu \right \rgroup \mkern-5mu \begin{array}{cccc} \mkern-12mu \\\mkern-12mu \\ \mkern-12mu \\ \mkern-12mu \end{array} \right),
\end{align} 
where the various entities are listed and briefly explained: 
\begin{itemize}
\item 
The iteration of the map $f_{a,b}$ on an initial condition gives the autonomous dynamics of the node. Some Neuron models like that of Fitzhugh-Nagumo (e.g., \cite{fitzhugh,rocsoreanu2012fitzhugh}) permit an individual neuron to have a regenerative self-excitation via a feedback term. Based on this principle, we  consider a highly simplistic discrete time model of the node dynamics where feedback from the node dynamics can cause self-excitation, and set
\begin{equation} \label{eq_isolated} f_{a,b}(x) := a \: x^3 - b \: x,
\end{equation}
 where parameters $0<a<1$ and 
$0<b<3$ (for a graph of $f_{a,b}$, see Fig.~\ref{figa}). Explanation on how a
parameter can cause self-excitation dynamics is made in Section~\ref{sub_sec_Isolated}.

\item  Adopting the principle that the individual drive from a connecting node has a larger magnitude if the incoming connection weight and the magnitude of the state value of the connecting node are together larger, we set the total sum of the drives from the other $(N-1)$ nodes into the node $i$ to be
$=$
\begin{equation} \label{eq_drive} - \: \sum_{j\not=i} \log \Big( \phi(W_{ij}(n)) \cdot (d + x_j(n))\Big),\end{equation} where  $W_{ij}(n)$ is the weight of the incoming connection from the $j^{\mbox{th}}$ node to the $i^{\mbox{th}}$ node;  $d >0$ is an offset parameter to enable a positive argument for the $\log$ function; $\phi : [0,\infty) \to [0,\infty)$ is an increasing function and is defined by
\begin{equation} \label{eq_phi}
\phi(\star) := \frac{1}{d} \; \; e^{\sqrt{\log(1 + \star)}}. 
\end{equation}
The function $\phi$ although increasing, has a monotonically decreasing slope (see Fig.~\ref{figc}). The net effect of the function $\phi$ can be thought of as the node's mechanism to have a relatively stronger inhibition to a stronger incoming connection -- node's internal mechanism to resist seizures by inhibiting stronger exogenous drive.  Lastly, the negative sign in \eqref{eq_drive} is only for mathematical convenience and has no physical significance. When the drive is zero, the dynamics of the node $i$ is in autonomous mode. 

\item The saturation function henceforth would be the mapping 
$\sigma : \mathbb{R} \to \mathbb{R}$ which is linear in a certain adjustable range, and its slope becomes zero asymptotically (see Fig.~\ref{figb}) defined by:
\begin{equation} \label{eq_sigma}
        \sigma(y) = \begin{cases}
                        y & \text{: $y \in [-p,p]$} \\
                        \frac{1}{r} \log(\frac{1}{r} + (y-p)) + k & \text{: $y > p$}\\
-\frac{1}{r} \log(\frac{1}{r} - (y+p)) - k    & \text{: $y < p$},                     
                    \end{cases}
\end{equation}
where 
$p>0$ and $\sigma$ is the identify function on 
$[-p,p]$;   $r>1$ determines the asymptotic rate at which the slope of $\sigma$ decreases and  $k = p - \frac{1}{r}\log(\frac{1}{r})$. (see Fig.~\ref{figb}); $\sigma$ is linear in the region $(-p,p)$ with a slope $1$, and the slope of $\sigma(y)$ approaches $0$ as $y\to \infty$ monotonically according to $\frac{1}{1+r(y-p)}$.
\end{itemize} 

Putting together the above into \eqref{eq_sketchy}, the update equation at the $i^{\mbox{th}}$ node is modeled by:
\begin{equation} \label{eq_FullModel_New}
x_{i}(n+1) = \sigma \bigg( f_{a,b}(x_i(n)) -    \: \sum_{j\not= i} \log \Big( \phi\big(W_{ij}(n)\big) \cdot (d + x_j(n))\bigg).
\end{equation}

Note that in the model in \eqref{eq_FullModel_New}, the weights $W_{ij}(n)$ are unspecified, but they actually can be solved for given a ECoG/EEG time-series $\{x_n\}$ (topic of Section~\ref{sec_Weights}). In Section~\ref{sub_sec_Isolated} and \ref{sub_sec_nonauto}, we describe the dynamics of a node when it operates in autonomous node and in the nonautonomous mode. In Section~\ref{sub_sec_thm}, we state sufficient conditions on the parameters $a$,$b$ and $d$, to ensure state-forgetting property of any evolving network arising out of \eqref{eq_FullModel_New}. However, to understand the model, a few remarks are in order concerning the role of 
the offset parameter $d$ and the saturation function $\sigma$ in \eqref{eq_FullModel_New}:

From \eqref{eq_drive}, it follows that the individual drive from the node $j$ into node $i$, at  time $n$ is given by $u_{ij}(n):= - \log \Big( \phi(W_{ij}(n)) \cdot (d + x_j(n))\Big)$. To make the role of the weights meaningful, a smaller weight $W_{ij}(n)$ should ensure smaller individual drive from node $j$, i.e., we must have $u_{ij}(n)$  close to zero whenever $W_{ij}(n)$ is close to zero.  We note that this would be the case whenever the offset parameter $d$ is sufficiently large.  To observe this, we remark on the behavior of $u_{ij}(n)$ as the weight $W_{ij}(n)$ gets close to zero. 
By definition of $\phi$ in \eqref{eq_phi}, when $W_{ij}(n)$ is close to zero, it follows that $\phi(W_{ij}(n))$ is close to  $\frac{1}{d}$. When $\phi(W_{ij}(n))$ is close to  $\frac{1}{d}$,  $u_{ij}(n)$ is close to $\log(\frac{d+x_j(n)}{d})$. For practical purposes we restrict $x_i(n) \in [-1,1]$ (see Section~\ref{sec_Weights}), and hence 
if $d$ is much larger than $x_j(n)$, then $\frac{d+x_j(n)}{d}$ is close to $1$, and $\log(\frac{d+x_j(n)}{d})$ is close to zero, and thus $u_{ij}(n)$ is very close to zero.  However, it is also to be noted that if $d$ is made unreasonably large, the signature of $x_j(n)$ on $u_{ij}(n)$ gets weaker, so $d$ is chosen in an appropriate range, although with a large value.  Physiologically, the parameter $d$ has no direct relation, but serves as a mathematical plug to overcome the difficulty in modeling the drive directly from the dynamical states of a connecting node that take both positive and negative values.   The offset $d$ has another role: its value could determine the state-forgetting property of an evolving network (see Theorem~\ref{Thm_Main}).

The particular choice of the saturation function $\sigma$ is not critical to the model; for the convenience of mathematical proofs, we need $\sigma$ to be differentiable and unbounded and hence its form. On the other hand, for  practical reasons, choosing such a $\sigma$, helps in the simulation stage to troubleshoot if each  of the parameters  $a$,$b$ and $d$ are chosen in an appropriate range lest the network lose the state-forgetting property. While the network does not have the state-forgetting property, the dynamics can be unbounded in the absence of a saturation function since the range of $f_{a,b}$ is unbounded and the numerical values on a computer could indicate an error.

\subsection{Isolated node-dynamics}\label{sub_sec_Isolated} 

We present a short summary of the dynamics of the family of maps $f_{a,b}$ described in \eqref{eq_isolated}. The detailed analysis of this family of maps is not the subject matter of this paper, and the discussion here is only intended to convey the point  that the dynamics of $f_{ab}$ adds notable features to the evolving network model. Recall that the parameters $a,b$ are restricted to $0<a<1$ and $0< b\le 3$. 
With $b\le 3$, it can be easily verified that there is an invariant closed interval $J$ symmetrically extending on either side of the real number $0$ given by $J := \big [-2\sqrt{\frac{b}{3a}} , \:\:\: 2\sqrt{\frac{b}{3a}} \big ]$, i.e., $f_{a,b}(J) = J$.  The dynamics of $f_{a,b}$ on $J$ is relevant to us since the dynamics outside $J$ escapes to $\pm \infty$. With $a$ fixed, and increasing $b$ through $0$ to $3$, the dynamics of this family of maps on $J$ takes a route to chaos qualitatively similar to the variation in $b$ from $0$ to $4$ in  the well studied logistic family of maps $L(x) = bx(1-x)$ on $[0,1]$ (for e.g., \cite{devaney1989}).
Elaborating this further, 
when $0< b <1$, the fixed point is globally stable, in the sense that every orbit asymptotically approaches the fixed point as $n$ tends to $\infty$; when $b = 1$, the fixed point loses stability and undergoes a period doubling bifurcation. As $b$ is increased beyond $2$, period doubling cascades occur alike in the logistic family of maps \cite{devaney1989}, and eventually there is a transition to chaos. Thus each node has the potential to have self-excitatory dynamics (oscillate or exhibit complicated behaviour) if the stability parameter for the node $b$ is set between $1 < b \le 3$.

\subsection{Driven node-dynamics} \label{sub_sec_nonauto}

Adding a time varying drive $u(n)$ to the update equation of $f_{a,b}$ by 

$$
x_i(n+1) = a \: (x_i(n))^3 - b x_i(n) - u(n),
$$
makes the mapping $x_i(n) \mapsto x_i(n+1)$ depend on $n$, and clearly the graph of such a mapping at any $n$ is a vertically shifted version of the graph of Fig.~\ref{figa} with the shift determined by $u(n)$. Since, we expect $u(n)$ to be changing with every $n$, the stable fixed point dynamics is not a feature of nonautonomous dynamics. 
As mentioned above, increasing $b$ beyond $1$, makes the dynamics move away from the fixed point at $0$. 
We make use of the fact that to move the nonautonomous dynamics away from the fixed point, it is not necessarily required to increase $b$ beyond $1$, but a sufficiently strong drive is capable of moving the dynamics. 
Also, some background observations and computer simulations suggest that larger the magnitude of drive, further away is the likelihood of the dynamics away from $0$.

\subsection{Is the evolving network state-forgetting?} \label{sub_sec_thm}

Having made some points on the nodal dynamics, we now state a result on the evolving network. Next, we state a theorem which says that when $d$ is sufficiently large, then any evolving network obtained from \eqref{eq_sigma} is state-forgetting. The complete proof  of this result is provided in Appendix~\ref{AppendixB}.

\begin{Theorem} \label{Thm_Main} 
Fix non-zero, positive reals $a,b,\lambda$ such that the following hold 

$$b \:<\: \lambda< 1 \:\:\: \:\:\:\& \:\:\: \:\:\:a \:< \: \frac{\lambda + b}{3}.$$ Also, fix a saturation function $\sigma$ as in Equation \eqref{eq_sigma} with $p>1$, and a time series $\{x(n) = (x_1(n),\ldots,x_N(n))\}$ such that $x_i(n) \in [-1,+1]$ for all $1 \le i \le N$ and all $n\in \mathbb{Z}$. 
There exists a real number $D>0$ such that for all $d>D$, if \eqref{eq_FullModel_New} holds for all $1 \le i \le N$ and $n \in \mathbb{Z}$  for some
$\{W(n)\} \subset \mathbb{R}^{N \times N}$, then the resultant evolving network is state-forgetting. 
\end{Theorem}

\section{Heuristic algorithm: directed weights from EEG/ECoG time-series} \label{sec_Weights}

Henceforth, for ease of jargon we call the $i^{\mbox{th}}$ component of a $N$-dimensional time series data $\{x(n)\}$,
i.e., $\{x_i(n)\}$ the data in the $i^{\mbox{th}}$ channel. Given $\{x(n)\}$, we present a method for solving the weights $W_{ij}(n)$ in\eqref{eq_FullModel_New} based on inferring a causal relationship between the nodes, which we call a synchrony measure.

To explain the methodology of solving weights from \eqref{eq_FullModel_New}, we rewrite \eqref{eq_FullModel_New} by rearranging as
\begin{equation} \label{eq_recast}
\sum_{j \not=i} \log(\phi(W_{ij}(n)) (d + x_j(n))) \: = \: a\left(x_i(n)\right)^3 - b \; x_i(n) - \sigma^{-1}\big(x_i(n+1)\big),
\end{equation}
where $\sigma^{-1}$ is the inverse of the saturation function $\sigma$.  Clearly, 
given a $N$-dimensional time-series $\{x(n)\}$ of EEG or ECoG clinical data, 
and having fixed $a$, $b$ in the model, the right hand side of \eqref{eq_recast}
can be computed easily for each $n$ from the $i^{\mbox{th}}$ channel data.  Let $r_i(n)$ denote the right hand side term in \eqref{eq_recast}. 

For a fixed $i$, we have a single equation 
\begin{equation} \label{eq_r_in}
\sum_{j \not=i} \log(\phi(W_{ij}(n)) (d + x_j(n))) \: = r_i(n),
\end{equation}
where there are $N-1$ unknown quantities $\left(W_{ij}(n)\right)$ as $j$ is varied from $1$ to $N$ with $j\not=i$. Hence, multiple solutions of $W_{ij}(n)$ could exist since we have more unknowns than equations. We regard seizure spread in the network due to sustained energy transmitted from a node that is seizing to the ones which is not. With this principle,  we hypothesize $W_{ij}(n)$ to be an increasing parametric function of a certain synchrony measure 
$\rho_{ij}(n)$ between the two distinct nodes $i$ and $j$. Physiologically, the synchrony measure between two nodes $i$ and $j$ represents the ease at which energy dissipates (a measure of conductance) from node $j$ into node $i$, and hence is directed -- nodes $i$ and $j$ could have different interconnections, and energy dissipation between nodes is not symmetric. 
To get a measure of such conductance, we let the synchrony measure $\rho_{ij}(n)$ depend on (i) a sustained influence from node $j$ into node $i$ measured by a leaky integrator placed between the two nodes 
(ii) the similarity between the average amplitude or power in the two channels up to time $n$.

The algorithm for computation of directed synchrony $\rho_{ij}(n)$ between two channels is described in 
Section~\ref{sub_sec_Alg}. 
We now explain the idea/reasoning behind some of the steps of the algorithm. The discussion can be read in conjunction with the various steps of the algorithm.

The very first step of the algorithm involves normalisation of the amplitude of data for setting the amplitude of the time-series in any channel lies between $[-1,1]$ (see (S1) of Section~\ref{sub_sec_Alg}).

In a succeeding stage of the algorithm, we compute the average power at each instant in a window of length $\tau$ in each channel
(see definition of $P_i(n)$ in (S6) of Section~\ref{sub_sec_Alg}). It turns out that the time interval in which there is sustained influence of one node on the other by the synchrony measure is determined by $\tau$. We comment on the choice of $\tau$ in Section~\ref{sub_sec_practical}

The directed synchrony $\rho_{ij}(n)$ depends on the product of two factors (see (S7) of Section~\ref{sub_sec_Alg}). One of the factors is the similarity between the powers in the $i^{\mbox{th}}$ and $j^{\mbox{th}}$ channels which we quantify by a similarity index $(1 - |P_i(n) - P_j(n)|)$.
Note that due to normalisation in (S1), the average power $P_i(n)$ in any  window is between $0$ and $1$, and hence $0 \le (1 - |P_i(n) - P_j(n)|) \le 1$, and the similarity index attains $1$ if and only if $P_i(n)$ and $P_j(n)$ are identical.

The other factor that contributes to  $\rho_{ij}(n)$ is $Q_{ij}(n)$, the output of a (directed) nonlinear leaky integrator placed between two distinct nodes and whose states are evaluated recursively (for details see (S8) to (S12) of Section~\ref{sub_sec_Alg})). The purpose of such a leaky integrator is to model the sustained influence of the signal strength in the node $j$ on node $i$ -- the word sustained points to the fact that the integrator leaks, and unless the influence is not sustained over a period of time, the influence diminishes. Any such leaky integrator in (S11) of Section~\ref{sub_sec_Alg} is a nonautonomous dynamical system in its own right, and hence its future states are obtained by iteration on the current state.   In essence, a leaky integrator sums up (or integrates) its previous state with what is fed to it typically as an input, but also leaks a small fraction of the sum while integrating. 
Here, we set the leaky integrator dynamics for the connection from node $j$ to $i$ as
\begin{align*} 
 \left \lgroup \mkern-5mu \begin{array}{ccc} Q_{ij}(n+1)), \mbox{ state} \\
 \mbox{of leaky integrator} \\ \mbox{at time } n+1 \end{array} \mkern-5mu \right \rgroup = 
& 
\left. \begin{array}{ccccc} \: \\ \mbox{leaky} \\ \mbox{integrating} \\ \mbox{function} \\ \:\end{array} \mkern-10mu \right (  \mkern-5mu
\left \lgroup \begin{array}{ccc} Q_{ij}(n), \mbox{state of} \\
 \mbox{leaky integrator }  \\ \mbox{at time  } n
 \end{array} \right \rgroup + 
\left. \left \lgroup \begin{array}{ccc} \mkern-5mu \bar{\bar{P}}_j(n), \mbox{ cumulative} \\ \mbox{average-fractional} \\
 \mbox{power at time } n  
 \end{array} \mkern-5mu \right \rgroup \mkern-5mu \begin{array}{ccccc} \mkern-12mu \\\mkern-12mu \\ \mkern-12mu \\ \mkern-12mu \\\mkern-12mu \end{array} \right),
\end{align*}
where the cumulative average power $\bar{\bar{P}}_j(n)$ in the $j^{\mbox{th}}$ channel is calculated  as in (S9) of the algorithm. This cumulative average fractional power is determined by first calculating the fractional power in the $j^{\mbox{th}}$ channel (see (S8) of the algorithm). The cumulative average of this fractional power is calculated as in (S9) of the algorithm; this cumulative average is calculated iteratively using the elementary idea: to calculate the average of a collection of $M$ numbers $(s_1,\ldots,s_M)$, if the average of $M-1$ numbers is given to be $A_{M-1}$, then the average of $M$ numbers $A_M$ can be updated by  $A_M = (M-1) A_{M-1}  + s_M$.

Once the synchrony $\rho_{ij}(n)$ from the time series from the algorithm in Section~\ref{sub_sec_Alg} is computed, then for convenience, we proceed to deduce the intermediate quantity $V_{ij}(n) := \phi(W_{ij}(n))$ from which the actual weight $W_{ij}(n)$ can be readily obtained (see \eqref{eq_Wij}).

We hypothesise $V_{ij}(n)$ to be an increasing parametric function of 
$\rho_{ij}(n)$ and is in the form  
\begin{equation} \label{eq_Vij}
V_{ij}(n) = \frac{1}{d}e^{\alpha_i(n) \cdot \rho_{ij}(n)},
\end{equation} where the parameter
$\alpha_i(n)$ can be solved as explained below. 
Substituting 
$V_{ij}(n) = \frac{1}{d}e^{\alpha_i(n) \cdot \rho_{ij}(n)}$ in Eq. \eqref{eq_r_in},  we have the following straightforward deduction to get an expression for $\alpha_i(n)$:
\begin{eqnarray} 
\sum_{j \not=i} \log\left ( \left(e^{\alpha_i(n) \cdot \rho_{ij}(n)}\right) \cdot  \left(\frac{d + x_j(n)}{d}\right) \right) &=& r_i(n), \nonumber \\
\sum_{j \not=i} \alpha_i(n) \cdot \rho_{ij}(n) &=& r_i(n) - \sum_{j \not=i} \log \left(\frac{d + x_j(n)}{d}\right), \nonumber \\
\alpha_i(n) & = & \frac{r_i(n) -\sum_{j \not=i} \log\left(\frac{d + x_j(n)}{d}\right)}{\sum_{j\not=i}\rho_{ij}(n)}.  \label{eq_alpha}
\end{eqnarray}

Once $V_{ij}(n)$ is found, the weight $W_{ij}(n)$ can be easily found since $\phi$ is invertible. From \eqref{eq_phi}, we get
\begin{equation} \label{eq_Wij} W_{ij}(n) = e^{\log^2(d\cdot V_{ij}(n))} -1. \end{equation}

\subsection{Algorithm to find the synchrony $\rho_{ij}(n)$} \label{sub_sec_Alg}

The algorithm is described in the following ordered steps (S1) to (S12).
\renewcommand{\labelenumi}{(S\arabic{enumi}).}
\begin{enumerate}

\item Normalize or scale the time-series data $\{x(n)\}$, by dividing each component in each channel by $\displaystyle{\max_{j,n}}(|x_j(n)|)$ so that upon normalisation the time-series satisfies: (i) every value of the time series in each channel lies between $[-1,1]$  (ii) the maximum of absolute value of the normalised time-series $\{x(n)\}$ (retaining the same notation as the original time-series) is $1$, i.e.,  $\displaystyle{\max_{j,n}}(|x_j(n)|)= 1$.

\item
Fix the parameters $a$, $b$ and $d$ of the network, and a large positive integer $\tau$. It is suggestive that the parameters $a$ and $b$ are chosen to satisfy $0.001 \le a \le 0.01$,  $ 1/3 \le b < 1$ and  $5000 \le \tau \le 10000$.  Once these values are fixed, (by Theorem~\ref{Thm_Main}) the
 value of the parameter $d$ determines the state-forgetting property of the evolving network. The state-forgetting property can be cross-verified after simulations (see Figure~\ref{fig: Network_Inf1}); the minimum value of $d$ required increases with the number of nodes in the network. For instance, in our simulation results in Section~\ref{Sec_results}, with $N=48$, the value of $d=500$ is found to be sufficient for the network to satisfy the state-forgetting property when $a=0.01$, $b=0.9$ and $\tau=5000$.

\item For all time $n<\tau$, initialise the cumulative average power in each of the $j^{\mbox{th}}$ channel to be $\bar{\bar{P}}_j(n) = 0$.

\item At time $n = \tau$, initialize, the leaky integrator state $Q_{ij}(n)$ a random number $(0,1)$ for all $i=1,\ldots,N$. 

\item Set a counter to an initial state, $\mathfrak{count}=0$ (to aid a calculation in (S9)).  

\item
Obtain the average power in every  channel $j$ of the time series at the time instant $n$ using the data in the window $[x_j(n-\tau), \ldots,  x_j(n)]$ 
as 
$\displaystyle{P_j(n) := \frac{1}{\tau+1} \sum_{k=0}^{\tau} |x_j(n-k)|^2.}$ (Note: Clearly, since the data is normalized in (S1), $\max(P_j(n)) = 1$)

\item Define the directed synchrony
$\displaystyle{\rho_{ij}(n) : = Q_{ij}(n) \cdot (1 - |P_i(n) - P_j(n)|)}$ for all $i$, $j$ running from $1,\ldots,N$ with $i\not= j$.

\item Find the fraction of the power in each channel $j$, using 
${\displaystyle\bar{P}_j(n) := \frac{P_j(n)}{\sum_{k=1}^N P_k(n)}}.$

\item For each $j$, compute the cumulative average $\bar{\bar{P}}_j(n)$ of the fractional power from  $\bar{\bar{P}}_j(n-1)$ and $\bar{P}_j(n)$ using 
$\displaystyle{\bar{\bar{P}}_j(n): = \frac{\mathfrak{count} \cdot \bar{\bar{P}}_j(n-1)+ \bar{P}_j(n)}{(\mathfrak{count}+1)}.}$

\item Denote $\mathfrak{t} : = Q_{ij}(n) + \bar{\bar{P}}_j(n)$, the quantity  that is to be fed into the leaky integrator to obtain its next state.

\item The $\tanh$ function when shifted up and to the right is a positive increasing map on $\mathbb{R}$ with a large slope in a very short interval; we use such a function to update
the leaky integrator state according to
$\displaystyle{Q_{ij}(n+1) := \frac{1}{2} \left(\tanh\left(4\mathfrak{t} -\frac{8}{5}\right) +1\right).}$

\item If $\rho_{ij}(n+1)$ needs to be computed, set $n=n+1$; $\mathfrak{count} = \mathfrak{count}+1$, and return to  (S6).~$\blacksquare$

\end{enumerate}

We recapitulate from Section~\ref{sec_SFN}, that only a state-forgetting network can attract all other iterates towards a unique entire solution. As a numerical verification of this, in Section~\ref{Sec_results} we show the ability of the evolving network model to generate a time series from an arbitrary initial condition that begins to resemble the original one in a very few time steps.

\section{Network Inference \& Potential Utility } \label{Sec_results}

We infer cortical connectivity from a 58 year old male patient diagnosed at Mayo Clinic. The patient had intractable focal seizures (seizures that fail to come under control due to treatment) with secondary generalization (seizures that start in one area and then spread to both sides of the brain). The patient underwent surgical treatment and pre-surgical evaluations were performed using a combination of history, physical exam, neuroimaging, and  electrographic recordings. For delineating the epileptogenic tissues, electrocorticography (ECoG) recording was performed at a sampling frequency of 500 Hz for four days continuously with grid, strip, and depth electrodes implanted intracranially on the frontal, temporal and inter-hemispheric cortical regions. A total of nine seizures were captured during the recording period and complete seizure freedom was achieved (ILAE class I) following the resection of pathological tissues in the left medial frontal lobe and medial temporal structures. All recordings were performed conforming to the ethical guidelines and under the protocols  monitored by the Institutional Review Board (IRB) according to National Institutes of Health (NIH) guidelines. The clinical data and ECoG recordings for this patient are publicly available on the IEEG portal (https://www.ieeg.org) with dataset ID `Study 038'.

\begin{figure}[!ht]
\centering
\includegraphics[scale=0.7]{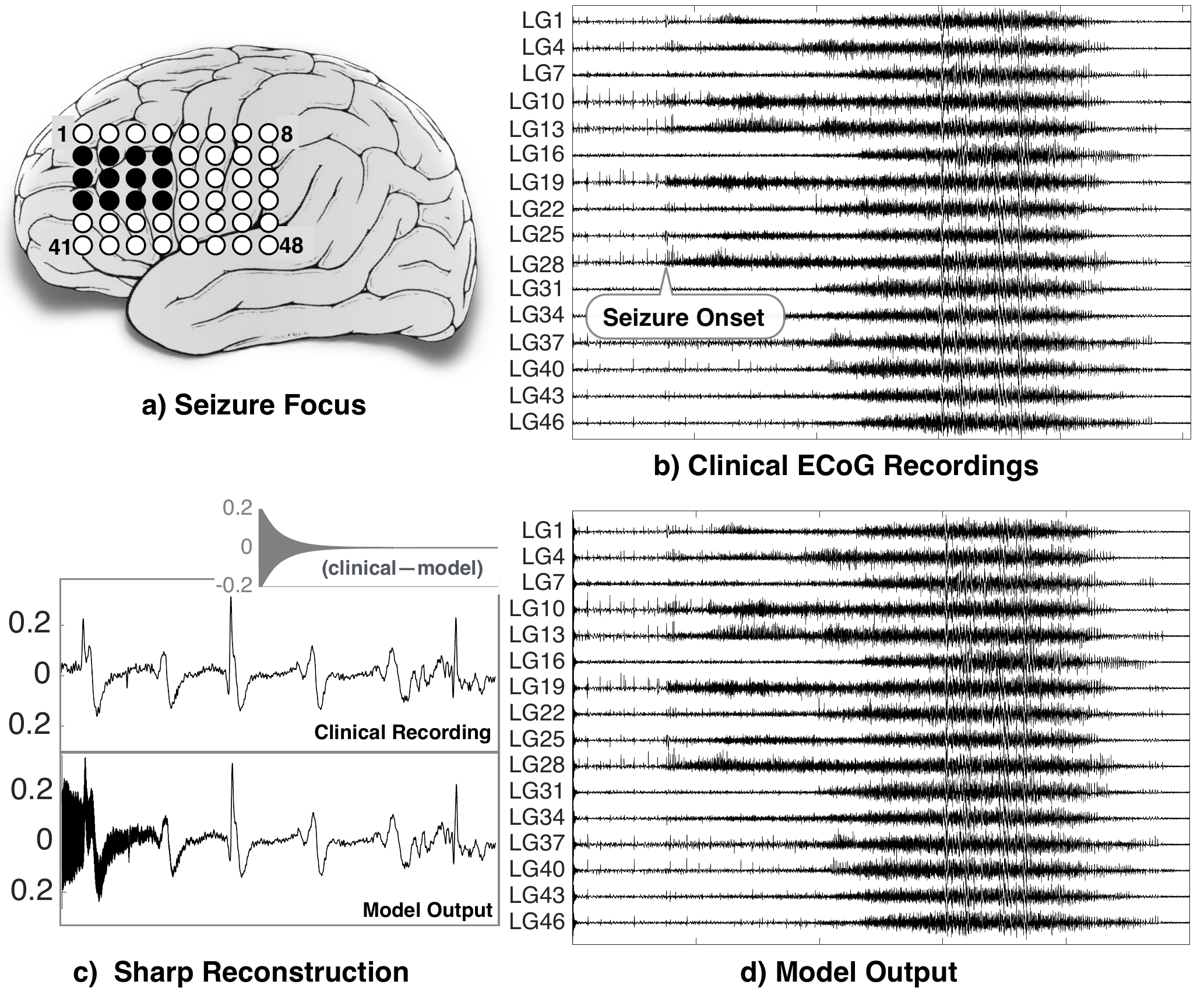}
\caption{{\bf Model simulations and sharp reconstruction of ECoG recordings} (a) shows the placement of electrodes in $(8 \times 6)$ grid on the brain schematic of a patient -- electrodes within clinical onset zone are shaded in black; (b) shows the time series of ECoG recordings from a few exemplary electrodes (channels) labeled with a prefix LG; (d)
shows the model generated time-series with an arbitrary initial condition in those channels shown in (b); (c) zooms on one of the channels in (d) and over the first 2000 samples to indicate the convergence of the model output to the ECoG recording from the model.} \label{fig: Network_Inf1}
\end{figure}

In our analysis, we consider only the electrodes in the $(8 \times 6)$ grid placed on the frontal lobe where both the seizure onset and its spread were captured. The electrodes which circumscribed the epileptogenic tissues are shaded in black on the schematic shown in Fig. \ref{fig: Network_Inf1}(a). A small segment of ECoG recording is sufficient to demonstrate the efficacy of our modelling framework. Therefore, we select an ECoG segment comprising of both pre-ictal and ictal activities to explore how the network evolves during seizures. We remove the noise artefacts and the power-line interference from the raw ECoG signals by applying a bandpass filter between $1-70$Hz and a notch filter at $60$Hz. The ECoG signals from a few exemplary electrodes (labeled with a prefix LG) are shown in Fig.~\ref{fig: Network_Inf1}(b). 

Next, we infer an evolving network synthesized from the ECoG data. We consider the cortical area under each ECoG electrode as a network node and apply algorithm detailed in Section~\ref{sub_sec_Alg} to get the directed and weighted edges or interconnections between the nodes so that \eqref{eq_FullModel_New} is satisfied. In particular with parameters $a=1$, $b=0.9$, $d=500$, $\tau=5000$ we apply the algorithm to obtain the weights of the interconnections between the nodes at each time sample of the data. To cross validate that the weights obtained eventually render a state-forgetting evolving network, we adopt the following procedure. With
an arbitrary initial condition $\hat{x}(n_{0})$ and the obtained weights at time $n_0$, we use \eqref{eq_FullModel_New} 
to obtain $\hat{x}(n_0+1)$. Repeating this, we obtain $\hat{x}(n_0+2)$, $\hat{x}(n_0+3)$ and so on.  The time-series in a few channels obtained from this procedure is shown in  Fig.~\ref{fig: Network_Inf1}(d). Apparently, the time-series in Fig.~\ref{fig: Network_Inf1}(b) and Fig.~\ref{fig: Network_Inf1}(d) are similar. Actually, the similarity is true barring a few hundred samples at the beginning (see Fig.~\ref{fig: Network_Inf1}(c)). This convergence of the model output from an arbitrary initial condition to the given time-series verifies state-forgetting property and validates our modelling framework for the study of network evolution during seizure onset and seizures. 

As a passing remark to the reader, the
model's ability to sharply reconstruct the given data
as indicated in Fig.~\ref{fig: Network_Inf1}(c) (plots of the first 2000 samples of an ECoG channel data, corresponding model output and their difference) is robust even for a different choice of parameters in the algorithm. In fact, Theorem~\ref{Thm_Main} tells us that such a reconstruction is possible as long as the parameters are in a certain range.

With an objective of identifying the nodes crucial for the seizure genesis, we study the network properties before and after the seizure onset. Consequently, we divide the ECoG recordings into two segments: pre-ictal (before seizure onset) and ictal (after seizure onset).

For each segment of this data, we compute an ensemble average of node strengths as explained next. It may be recalled from Section~\ref{sec_Model} that the drive from node $j$ to $i$ at time $n$ is given by $u_{ij}(n):= - \log \Big( \phi(W_{ij}(n)) \cdot (d + x_j(n))\Big)$. For each node $i$, we compute its mean incoming drive strength of node  over a time period $T$ as $\displaystyle{\frac{1}{|T|} \sum_{n \in T} \frac{1}{N-1}\sum_{j=1,j\not=i}^N |u_{ij}(n)|,}$ where $N$ is the number of nodes, and $|T|$ is the number of sample points in $T$.  In a similar vein, we compute the mean outgoing drive  strength of node $i$ over a time period $T$ as 
$\displaystyle{\frac{1}{|T|} \sum_{n\in T} \frac{1}{N-1}\sum_{j=1,j\not=i}^N |u_{ji}(n)|.}$

\begin{figure}[!ht]
\centering
\includegraphics[scale=0.75]{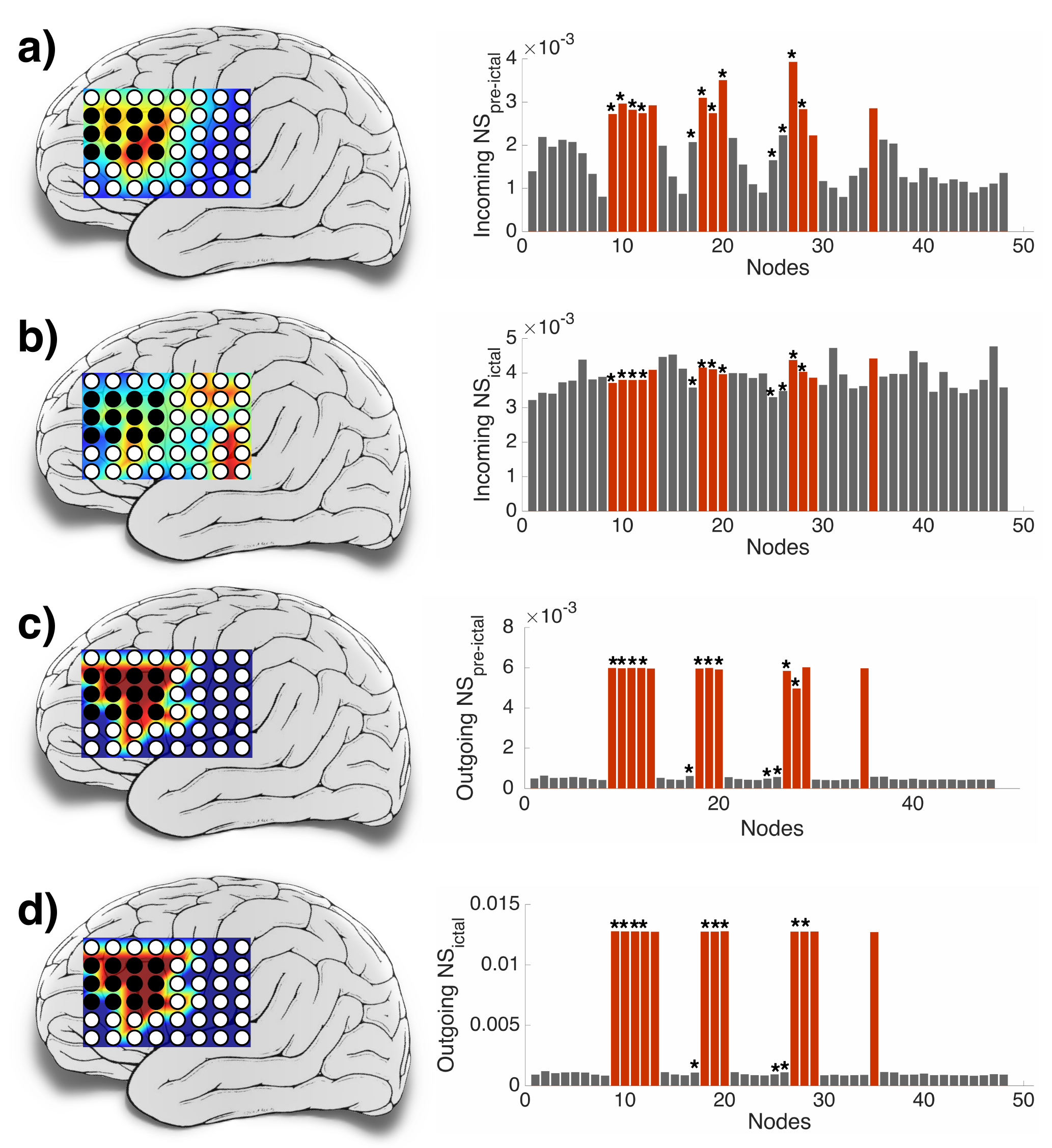}
\caption{ Illustration of the variation in the mean incoming and outgoing node strengths across nodes during the pre-ictal and ictal period (the word `mean' is omitted in the labels only for a shortage of space). Nodes shaded in black are from the clinically delineated seizure focus and are also depicted by asterisks on the corresponding columns in the bar plots. Nodes whose node strengths are shown as columns in red
correspond to the theoretical seizure focus identified from the model simulations. The color code on the brain schematic at the nodes is for convenient viewing and indicates the corresponding relative node strength in the bar plot, but only approximately} \label{fig: Segmentation_EEG}
\end{figure}

Figure \ref{fig: Segmentation_EEG} shows the mean incoming and outgoing node strengths for each node during the pre-ictal and ictal periods. Fig. \ref{fig: Segmentation_EEG} (a) indicates a strong correlation of the mean incoming node strengths with the clinical seizure onset zone or seizure focus (depicted by the electrodes shaded in black) during the pre-ictal period, while Fig. \ref{fig: Segmentation_EEG} (b) indicates a more  uniform spread of mean incoming node strengths across all nodes during the ictal period. The correlation of  
the mean incoming node strengths with the clinical seizure onset zone during the pre-ictal period can be understood as that the external drives from other nodes play a supportive role in initiating seizures (a hypothesis of our model as well). Since during the ictal period all channels seem to have similar average power due to seizures, all nodes are likely to receive a similar quantum of external drive. Hence, the mean incoming drive strength appears to be uniform.

In contrast, Fig.~\ref{fig: Segmentation_EEG} (c, d) shows that the distinctively higher mean outgoing node strength (during both  ictal and pre-ictal period) is much strongly ``correlated" with the clinical seizure onset zone than that shown by the corresponding mean incoming node strengths. Therefore, from Fig.~\ref{fig: Segmentation_EEG} (c, d),  it is evident that the nodes whose node strengths are shown as columns in red (i.e., nodes ${9, 10, 11, 12, 13, 18, 19, 20, 27, 28, 29, 35}$) are the outgoing hubs. We refer to this set of nodes as the theoretical seizure onset zone.

Incidentally in this experiment there are 12 nodes in the theoretical seizure onset zone, similar in cardinality to the set of the nodes at the site of clinical resection ( i.e., nodes $(9, 10, 11, 12, 17, 18, 19, 20, 25, 26, 27, 28)$). We refer to the nodes at the site of clinical resection as the clinical seizure onset zone. It is evident that the theoretical seizure onset zone and the clinical seizure onset zone point to very similar regions in the brain. Thus, the model has the potential utility of identifying an approximate seizure focus zone prior to surgery by analysing some pre-ictal  and ictal data.

Although the theoretical and clinical onset zones differ slightly, from a  theoretical standpoint of view, we can hypothesise the outcome of  the surgical removal of these two zones and compare these outcomes if we have a suitable objective measure. To explain this, 
let $\mathfrak{R}$ be the indices of the subset of the nodes that are omitted from a network of $N$ nodes in a simulation experiment, i.e, the nodes in $\mathfrak{R}$ are made non-functional by setting $W_{ij}(n) = 0$ for all $n$ if either $i$ or $j$ belongs to $\mathfrak{R}$. Using such modified weight matrices in \eqref{eq_FullModel_New} we can obtain a time-series $\{y(n)\}$ of the depleted network can be calculated by starting with any initial condition that satisfies $y_i(n_{0}) = 0$ whenever $i \in \mathfrak{R}$. In other words, the signal has to have zero amplitude in all the channels corresponding to the nodes in $\mathfrak{R}$ since they are assumed to not participate in the network dynamics.

We quantify the efficacy of the removal of the nodes that belong to $\mathfrak{R}$ by calculating the ratio of the average power in those channels outside $\mathfrak{R}$ prior to removal of nodes and after it by:
$$
P_{\mbox{original}} = \frac{1}{|T|} \sum_{n\in T} \: \: \left(\frac{1}{N-|\mathfrak{R}|} \sum_{j\in \mathfrak{R}} (x_j(n))^2\right),
$$ where  $|T|$ denotes the number of samples over a time-interval $T$ and $|\mathfrak{R}|$ denotes the cardinality of the set $\mathfrak{R}$. 
Similarly the
average power in the signal generated from the depleted network can be calculated by
$\displaystyle{
P_{\mbox{depleted}} = \frac{1}{|T|} \sum_{n\in T} \: \: \left(\frac{1}{N-|\mathfrak{R}|} \sum_{j\in \mathfrak{R}} (y_j(n))^2\right)},
$
and we calculate the efficacy of removing the nodes $\mathfrak{R}$ by the ratio $G(\mathfrak{R}) = P_{\mbox{original}}/P_{\mbox{depleted}}.$ 

We tabulate the values of $G(\mathfrak{R})$ below, where $\mathfrak{R}$ is determined by the following choices: (i) the clinical seizure onset zone (ii) theoretical seizure onset zone (outgoing hubs) (iii) an arbitrary set of 12 nodes (iv) 12 nodes randomly drawn (average of $G(\mathfrak{R})$ is tabulated in this case).  Now, we discuss the potential application of the model in a clinical study when surgery is part of the treatment. The aim of a clinical surgery would be to a minimise the removal of tissue from the brain along with subsiding seizures post surgery.  In our study above, the removal of a set of nodes $\mathfrak{R}$ from the evolving network is intended to simulate a surgical removal of the tissue beneath the nodes in $\mathfrak{R}$. A time series $\{y(n)\}$ from the depleted network is intended to simulate the resultant clinical ECoG time series observed after surgery, and the quantity $G(\mathfrak{R})$ is meant to measure the efficacy of the surgery. Thus prior to an actual surgery, the theoretical efficacy of surgical removal of nodes can be studied by using the corresponding depleted network. From the tabulated values of $G(\mathfrak{R})$, we observe that the removal of nodes from the outgoing hubs that form the theoretical onset zone maximises the efficacy of a surgery when (the tissue beneath) a set with 12 nodes is assumed to be removed in each simulation. In fact the large value of $G(\mathfrak{R})$ obtained after the removal of the nodes in the theoretical onset zone translates to complete subsiding of seizures in the resultant time-series (not shown here) generated from the depleted network. When a small subset of the outgoing hubs were retained in the network, $G(\mathfrak{R})$ was still observed to be smaller and the seizures continued to persist from the resultant time-series (not shown here) generated from the depleted networks. From this ability of outgoing hubs to subside seizures in the rest of the network upon their omission, we also refer the outgoing hubs to as spreaders. For a reader interested in finding an optimal set of nodes for subsiding seizures, we note that one does not have to necessarily aim to get a very large value of $G(\mathfrak{R})$, but any $G(\mathfrak{R})$ that yields a seizure free time-series is adequate.

\begin{center}
\begin{tabular}[c]{|p{3.5cm}||p{8cm}|p{1.5cm}|}
\hline
Onset zone & $\mathfrak{R}=$ & $G(\mathfrak{R})$  \\ \hline \hline
Clinical & $(9, 10, 11, 12, 17, 18, 19, 20, 25, 26, 27, 28)$ & $7.20$ \\  \hline
Theoretical & $(9, 10, 11, 12, 13, 18, 19, 20, 27, 28, 29, 35)$ & $27.24$ \\ \hline
Random &  $(6,7,11,13,24,25,34,38,40,43,45,48)$ & $1.51$ \\ \hline
20 sets of Random  & each trial (not listed here) has 12 nodes & 
$1.69$ \\
\hline
\end{tabular}
\end{center}

We remark that in our simulations, the mean incoming (outgoing) node strengths and the mean incoming (outgoing) weights of a node exhibit similar variation across nodes, but we employ the former as it accounts for more accurate interaction between two nodes since it subsumes both the effect of the weight of the interconnection and the signal strength of the node that connects.

\subsection{Practical issues and modifications}  \label{sub_sec_practical}

The parameter $\tau$ determines the length of the windows in which average powers are compared (in the algorithm in Section~\ref{sub_sec_Alg}) while computing the synchrony measure. Since the power during the inter-ictal period (period between two distinct seizures) is negligible in all channels and seizures are relatively short epochs, if $\tau$ is made extremely large then the synchrony measures would not show distinct variation across nodes,  and we could end up with no useful network information as the weights lose physical significance. It is required to set $\tau$ in a range so that  some noticeable  pre-ictal activity or larger power  in the data that typically occurs (e.g.,\cite{khosravani2009spatial}) just prior to seizures is captured in the synchrony measure. In this paper, we set $\tau = 5000$ so that with a sampling rate of $500$ Hz we use about 10 seconds of data for such a power comparison. However, it turns out that the mean node strengths show similar variation even if $\tau$ is increased in a range of up to $15000$, but a large $\tau$ increases computational complexity.

\begin{figure}[h]
\label{figicoc}
\centering
\subfigure[Average incoming weights]{\includegraphics[width=8cm,height=6cm]{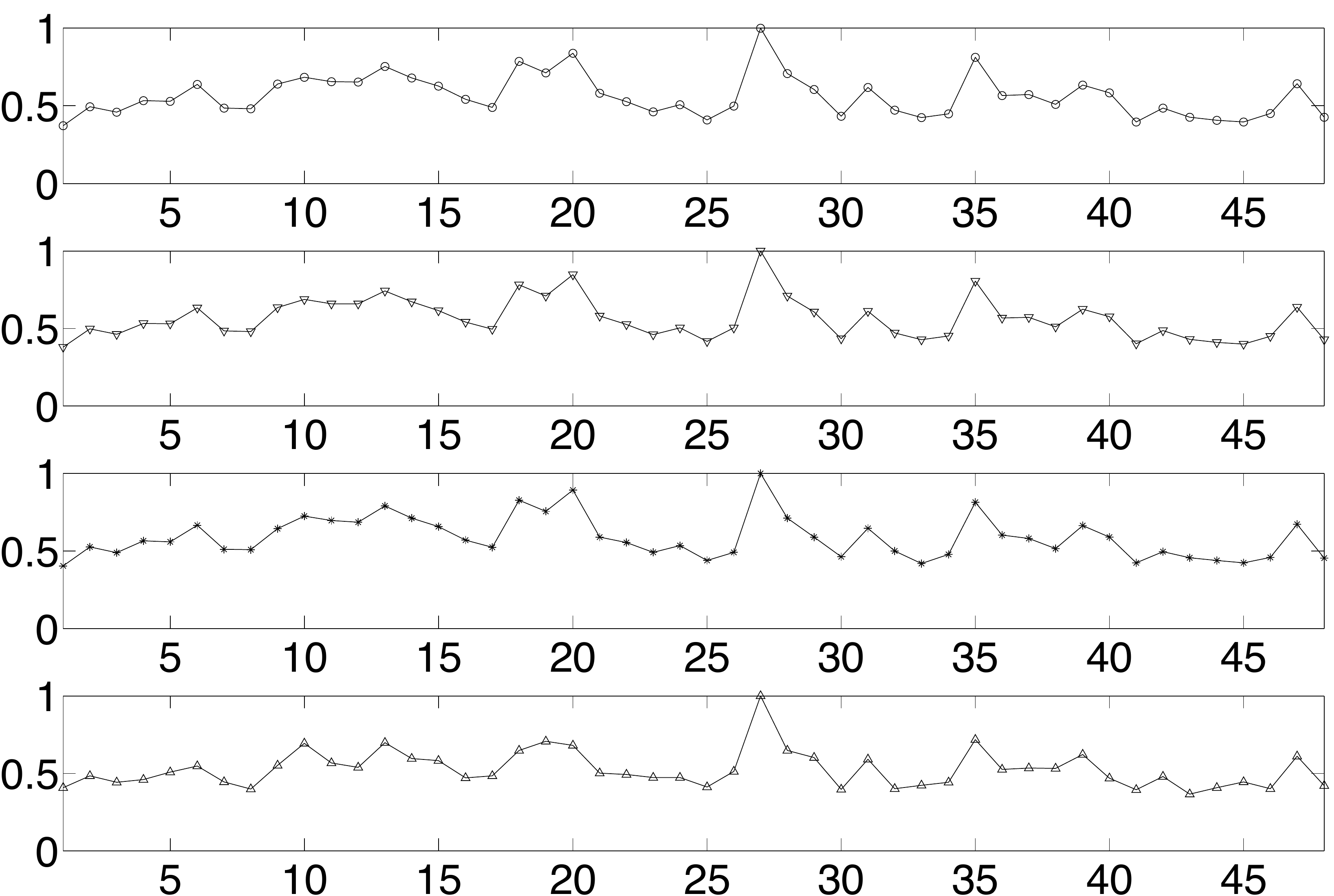}\label{figic}}
\centering
\subfigure[Average outgoing weights]{\includegraphics[width=8cm,height=6cm]{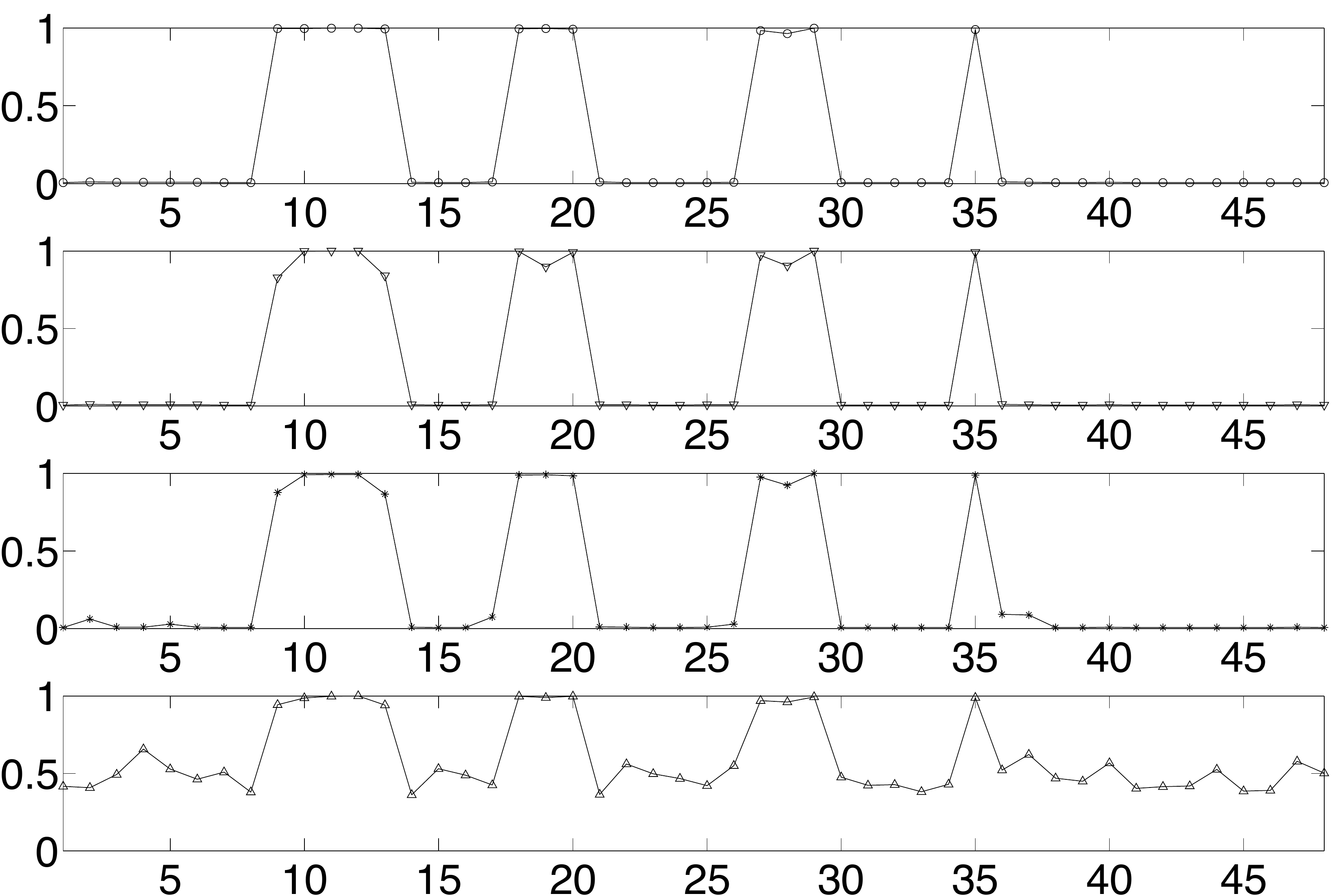}\label{figoc}}
\caption{Nodes Vs. Normalised average incoming/outgoing weights (graph connected by a line) computed with/without additive uniform iid noise; Without noise in Panel 1; Panel 2, 3, 4 use respectively uniformly distributed noise in 
$[-0.1,0.1]$,  $[-0.25,0.25]$ and $[-0.5,+0.5]$.}
\end{figure}

In conventional machine learning problems, complicated models are avoided fearing an over-fitted model. An over-fitted network model is usually a static network with a very large number of nodes and it tends to model the random noise or anomalies in the data rather than the underlying relationship. We evaluate the sensitivity of the weight computation algorithm or the model  by analysing the change in computed weights with additive artefact noise to the data; we use weights rather than  node strengths since we are analysing the model sensitivity and not evaluating a seizure focus.   By adding realizations of zero mean iid noise having uniform distribution to the above analysed patient's clinical ECoG data (taking values in $[-1,1]$), we plot in Fig.~\ref{figic} and Fig.~\ref{figoc} the normalised average incoming and outgoing connections for different noise variances -- the average weights are normalised to have a maximum average value of $1$ for the purpose of comparison across different noise variances.   The pattern in the variation of the average weights has many similar features even when the noise variance is high (fourth panel in Fig.~\ref{figic} and Fig.~\ref{figoc}).  With such a high variance of noise, we observe from Fig.~\ref{figoc} that all the nodes in the theoretically determined focus are still outgoing hubs. Similar robustness of  the pattern in normalised weights were observed when the data was quantised or when the parameters ($a$, $b$ and $d$) in our algorithm were varied.

Lastly, we remark that the model is flexible to changes and can be modified to work  as long as the state-forgetting property is satisfied. In particular, the dynamical system at each node in \eqref{eq_isolated}  and inferring causality in the Algorithm in \ref{sub_sec_Alg} can be made more sophisticated or changed to be based on a deeper or alternate physiological understanding of how epileptic seizures are generated.

\section{Conclusions}
\label{sec_conclusion}

In this paper, we have presented a modeling approach to transform data into an evolving network that has the state forgetting property or robustness to initial conditions. Although, this principle of transformation can be applied to data originating from any source, the particular evolving network model concerned an epileptic network. 
Although time-series can be modeled with reasonable accuracy by static networks of very large dimension such as recurrent neural networks, the individual nodes in such case have no physical significance and they do not correspond to the regions in the brain. The interconnection-strengths between the nodes were solved via a causal relationship inferred from the sustained power similarity between two nodes.

Since the model can produce a sharp reconstruction of the time-series, there is flexibility to study the interconnections in the evolving network in conjunction with the value of the time series at any time point, or over a time-interval. We have observed from modeling clinical data of an epileptic patient that certain nodes turn into outgoing hubs during the pre-ictal period and persist through the ictal period. These outgoing hubs also have a location in the brain similar to that of the clinically identified seizure onset zone (seizure focus). This hints that the (pathological) seizure causing areas in the brain could be having strong outgoing connections during seizure onset and during seizures.

Further, to quantify the influence of such outgoing hubs on the network dynamics, we have presented a result quantifying the theoretical efficacy of removal of such nodes in subsiding seizures. Of course, the result is based on data of a single patient, but then a comprehensive study of the effect of a removal of any node in an evolving network has to be backed up with a justification -- since altering even a single interconnection could alter the entire network dynamics, a reliable model should show small changes in the network dynamics for small changes in the weights.  In a sequel paper, we present a mathematical justification for inferring network information from a depleted network where we establish that changing the interconnections results in a continuous change in the time series whenever the evolving network is state-forgetting.  Such a result is needed to justify a more detailed study of the theoretical efficacy of removal of nodes.   In fact, it is possible to prove a very strong result showing that the effect of the removal of a node can be studied reliably if and only if a `nontrivial' evolving network has state-forgetting property.

\section*{Acknowledgments}
The author would like to thank Nishant Sinha at Nanyang Technological University for assistance in presenting plots in Section~\ref{Sec_results} and some discussions there in.  The author thanks the team of the IEEG portal (https://www.ieeg.org) for providing access to the ECoG data. Access to the image(s) and/or data was supported by Award number U24NS06930 from the U.S. National Institute of Neurological Disorders and Stroke. The content of this publication/presentation is solely the responsibility of the author, and does not necessarily represent the official views of the U.S. National Institute for Neurological Disorders and Stroke or the U.S. National Institutes of Health. The author thanks an anonymous referee for pointing out several typographical errors.

\section*{References}
%\bibliography{references.bib}

\appendix
\renewcommand*{\thesection}{\Alph{section}}

\section{Appendix: Autonomous vs. Nonautonomous Dynamics}\label{AppendixAA}

When interconnections change with time in a network of dynamical systems, the overall dynamics of the entire network is nonautonomous -- autonomous dynamics is rendered when systems evolve according to a rule fixed for all time, and when it is not fixed (here, due to the time-varying interconnections), nonautonomous dynamics emerges (see Section~\ref{sec_SFN} for definitions). The evolution of an autonomous system depends on only the initial condition presented to it, whereas in a nonautonomous system the evolution depends both on the initial condition and the time at which it is presented. The evolution of an autonomous system in discrete-time can be described by an update equation $x_{n+1}=f(x_n)$, where $f$ is a function governing the rule for evolution. The evolution of a nonautonomous system can be described by an update equation $x_{n+1}=f_n(x_n)$, where a family of maps $\{f_n\}$ govern the rule of evolution.

We explain the advantage of considering a nonautonomous model. Consider a two-dimensional autonomous dynamical system  which updates its dynamics according to the rule $(x_{n+1},y_{n+1}) = (f(x_n,y_n), g(x_n,y_n))$ ($f$ and $g$ are any functions whose domain and range are $\mathbb{R}^2$ and $\mathbb{R}$) to generate a $2$-dimensional time series $\{(x_n,y_n)\}$. Suppose in reality, only the first coordinate or dimension of the time series, i.e,  $\{x_n\}$ is observable. Then, in general it is not possible to model the time-series $\{x_n\}$ as the dynamics of a one-dimensional dynamical system, in other words it is not possible to find a function (map) $h$ so that $x_{n+1} = h(x_n)$ holds for all $n$ since $x_{n+1}$ inherently depends on $y_n$. However, it is possible to find a nonautonomous system, i.e., a family of maps $\{h_n\}$ so that $x_{n+1} = h_n(x_n)$. In general, whenever we are unable to completely observe all the variables from a dynamical system (including nonautonomous systems) it is possible to synthesise a nonautonomous system that models the observed variables.  In real world systems, the non-autonomous system can account for obfuscated factors such as the unfeasibility of observing all variables, exogenous influence etc. by subsuming their net-effect by a collection or family of maps that vary with time.

\renewcommand{\theequation}{B-\arabic{equation}}
\counterwithin{Theorem}{section}

\section{Appendix: Formal Definition of State Forgetting Property}\label{AppendixA}

Mathematically, viewing an evolving network as a static network being acted upon by an exogenous stimulus that influences the network connections at each time step gives us the advantage of employing the formal framework of the echo-state static networks theory in machine learning \cite{jaeger2001,manjunath2013}. We recall some notions
\cite{jaeger2001,manjunath2013} and adapt to our setup. We begin now with a formal description.

Here, and throughout, only the Euclidean distance is employed as a metric/distance on any two elements of $\mathbb{R}^d$.  Also $I$ represents a nondegenerate compact interval of $\mathbb{R}$, and $I^N$ denotes the $N$-times product $(I \times \cdots \times I)$.

Following \cite{manjunath2013, manjunath2014} and several others (e.g.,  \cite{elaydi2005introduction,kloeden2013discrete,kloeden2011nonautonomous}), we define a \emph{discrete-time nonautonomous system} on a metric space $X$ as a
sequence of maps $\{g_n\}$, where each $g_n : X \to X$ is
a continuous map.  Any sequence $\{x_n\} \subset X$ that satisfies $x_{n+1} = g_{n}(x_{n})$ for all $n\in \mathbb{Z}$ is called an entire-solution of $\{g_n\}$.
For the discussion in this paper it is sufficient to consider an evolving network with each of the nodes having a one-dimensional dynamical system on an interval $I$, i.e., for an isolated node, there is some function mapping $I$ to $I$ as it is an autonomous system. We formally define an evolving network  below:
\begin{Definition} \rm \label{def_EN}
Let $I$ be a compact interval of $\mathbb{R}$, and let $N\ge 1$ be an integer. 
Suppose a function $G$ has a real-valued $N \times N$ matrix $W(n)$, and a value in $I^N$ as its arguments and maps it to a value in $I^N$ for each $n\in \mathbb{Z}$, i.e., $G : \mathbb{R}^{N \times N} \times I^N \to I^N$ is such that
\renewcommand{\labelenumi}{(\roman{enumi})}
\begin{enumerate}
\item $G$ is continuous and
\item $G$ is defined by a set of $N$ (continuous) coordinate maps $h_i: I^N \times \mathbb{R}^N \to I$ such that $h_i : (x_i(n), W_{\bar{i}}(n)) \mapsto x_i(n+1)$, where $W_{\bar{i}}(n) \in \mathbb{R}^N$ represents the $i^{\mbox{th}}$ row of a matrix $W(n)$, and  $x_i(n)$ the $i^{\mbox{th}}$ component of $x(n+1)$; here $h_i$ generates the node dynamics at the $i^{\mbox{th}}$ node,  
\end{enumerate}
then such a map $G$ is called a \textit{family of evolving network}.  If in particular, a bi-infinite sequence of matrices $\{W(n)\} \subset  \mathbb{R}^{N \times N}$ is fixed, then the sequence of mappings $g_{_{W(n)}}(\cdot) := G(W(n), \cdot)$ is referred to as an evolving network. 
\end{Definition}
Clearly, any evolving network $g_{_{W(n)}}(\cdot) := G(W(n), \cdot)$ is also a nonautonomous system $\{g_n\}$ on $I^N$ if we set $g_n := g_{_{W(n)}}$ for all $n$. We call
$W(n)$ the connectivity weight matrix or just the weight matrix.

We next define the state-forgetting property of a family of evolving networks and also for a particular evolving network.

\begin{Definition} \rm \label{def_SF}
Let $G: \mathbb{R}^{N \times N} \times I^N \to I^N$ be a family of evolving networks. If for each choice of $\{W(n)\}$, the evolving network 
$g_{_{W(n)}}(\cdot) := G(W(n), \cdot)$ has 
exactly one entire-solution, then the family of evolving networks $G$ is said to be state-forgetting.  In particular, for any fixed
 $\{W(n)\}$, the corresponding evolving network (or the nonautonomous system) $\{g(W(n),\cdot)\}$ is said to be \textit{state-forgetting} if it has exactly one entire-solution. 
\end{Definition}

\renewcommand{\theequation}{C-\arabic{equation}}
\section{Appendix: Proof of State Forgetting Property}\label{AppendixB}

We first recall some preliminaries and known results that help us in presenting the proof of Theorem~\ref{Thm_Main}.

It is convenient to have the following alternate representation of a nonautonomous system.
Let $\mathbb{Z}^2_{\ge}$ the collection of all integer tuples $(n,m)$ so that $n\ge m$, i.e., 
$\mathbb{Z}^2_{\ge} := \{(n,m) : n,m \in
\mathbb{Z} \: \&  \: n \ge m\}$.  
Given a nonautonomous system $\{g_n\}$ on $X$, it is convenient (see \cite{manjunath2013}) to denote the nonautonomous system through a function $\phi: \mathbb{Z}^2_{\ge} \times X \to X$ called a {\em process}  on $X$ that satisfies: $\phi(m,m,x) := x$ and $\phi(n,m,x) := g_{n-1}\circ \cdots \circ g_m(x).$  We say $\phi$ has the state-forgetting property if $\{g_n\}$ has the state-forgetting property. 

The following Lemma is reproduced from \cite[Lemma 2.2]{manjunath2013}.  
\begin{Lemma} \label{Lemma_backward_contraction} Let $\varphi$ be a
  process on a compact metric space $X$ metrized by $d_X$.
  Suppose that, for all $n \in \mathbb{Z}$, there exists a sequence of
  positive reals $\{\delta_j\}_{j=1}^\infty$ converging to $0$ such
  that $d_X(\varphi(n, n-j, x), \varphi(n ,n-j, y)) \le \delta_j \; d_X(x,y)$
  for all $x,y\in X$ and for all $j \in \mathbb{N}$. Then $\phi$ has the state-forgetting property, i.e., there is
  exactly one entire solution of the process $\varphi$.
\end{Lemma}

The essence of the following result is that if $\phi$ has the state-forgetting property then all initial conditions converge to the unique entire-solution in Proposition~\ref{prop_pullback}. For a detailed explanation see \cite{manjunath2014} and references therein. 

\begin{Proposition} \label{prop_pullback} Let $\varphi$ be a
process on a compact metric space $X$ metrized by $d_X$. Suppose $\{\vartheta_n\}$ is the unique entire-solution of $\varphi$, then 
$\lim_{j \to \infty} d_X(\varphi(n, n-j, x), \varphi(n ,n-j, \vartheta_n)) = 0$ for all $x$ and $n$.  
\end{Proposition}

We repeatedly use the following standard inequalities involving the infinity norm and spectral norm: $\| x \|_\infty \le \| x \|_2$ and $\| x \|_2 \le \sqrt{N} \| x \|_\infty$ for all $x\in \mathbb{R}^N$; when $A \in \mathbb{R}^{N \times N}$, 
$\| A \|_2 \le \sqrt{N} \| A \|_\infty$.

A generalization of the mean-value theorem in one-dimensional calculus to
higher dimensions results in the following mean-value inequality (e.g.,
\cite{furi1991}). This result can be stated as:
if $\varphi : V \to \mathbb{R}^n$ is a
$C^1$-function where $V$ is an open subset of $\mathbb{R}^n$ then for
any $x,y \in V$
\begin{equation} \label{ineq_MVT}
\|\varphi(x) - \varphi(y) \|_2 \le \sup \{ \|\varphi^{'}(z)\|_2 : z
\in V \} \cdot \|x-y\|_2,
\end{equation}
where $\|\cdot \|_2$ is the  $L^2$-norm, and $\|\varphi^{'}(z)\|_2 :=
\sup(\|\varphi^{'}(z) x\|_2 : \|x\|_2 = 1)$ is the induced norm of the
Jacobian of $\varphi(\cdot)$ at the point $z$.

To simplify the notations in the proof of Theorem~\ref{Thm_Main}, whenever $y =(y_1,\ldots,y_N)$, we adopt the notation
$\overline{\log}(y) := (\log(y^1), \log(y^2),
\ldots , \log(y^N))\transp$ and \\ $\overline{\sigma}(y) := (\sigma(y^1), \sigma(y^2),
\ldots , \sigma(y^N))\transp$, where $\overline{\log}(y)$ and $\overline{\sigma}(y)$ are column vectors and the superscript $(\cdot)\transp$ represents the transpose of the vector $(\cdot)$.

{\bf Proof of Theorem~\ref{Thm_Main}. }  Fix a $\{W(n)\} \subset \mathbb{R}^{N \times N}$ that satisfies \eqref{eq_FullModel_New} for all $1\le i \le N$ and $n\in \mathbb{Z}$. Let $\{V(n)\}$ be defined by $V_{ij}(n) = \phi(W_{ij}(n))$, where $\phi$ is as in \eqref{eq_phi}. Since fixing a $\{W(n)\}$ or a $\{V(n)\}$ ($V_{ij}(n)$ related to $W_{ij}(n)$ by \eqref{eq_phi}) amounts to fixing an evolving network, we have fixed a nonautonomous system.

In the notion of a process, the evolving network update obtained by varying $i$ from $1$ to $N$ in \eqref{eq_FullModel_New} can be represented at time $n$ by

$$\varphi(n,n-1,y) = \overline{\sigma}\bigg(F_{a,b}(y) - L_{n-1}(y)\bigg),$$ 
where $y\: \in \: [-1,1]^N$, $\: F_{a,b}(y_1,\ldots,y_N) := (f_{a,b}(y_1),\ldots,f_{a,b}(y_N))$ and  $\: L_{n-1}(y) := $ \\ $\overline{\log}\left(V(n-1) \cdot (d+y)\right)$, with $(d+y)$ as the column vector $(d+y_1, \ldots, d+y_N)$.  Since the function $\sigma$ in \eqref{eq_sigma}, ensures $\| \sigma(y_1) - \sigma(z_1) \| \le \| y_1 - z_1 \|$, we have 
\begin{equation} \label{ineq_sigma}
\| \overline{\sigma}(y) - \overline{\sigma}(z) \|_{\infty} \le \| y - z \|_{\infty} \mbox{ for all } y,z \in \mathbb{R}^N.
\end{equation} 
Since the choice of $\{W(n)\}$ was made arbitrarily, if we show that $\varphi$ has exactly one-entire solution then we would have shown that the evolving network is state-forgetting by Definition~\ref{def_SF}. Since $\phi$ is a process on $[-1,1]^N$, a compact subset of $\mathbb{R}^N$, we make use of Lemma~\ref{Lemma_backward_contraction} to show $\varphi$ has exactly one-entire solution. We employ the metric induced by the norm $\| \cdot \|_\infty$ for getting an inequality of the form in Lemma~\ref{Lemma_backward_contraction}.

Let $\delta$ be any real number in $(\lambda,1)$. We first aim to show that there exists a $D$ (that depends on $\delta$) such that for all $d>D$, the following inequality is satisfied. 
\begin{eqnarray} \label{eq_mainineq}
\| \varphi(n,n-1,y) - \varphi(n,n-1,z) \|_{\infty} & \le \delta \| y - z \|_{\infty}  \mbox{ for all } y,z, \in [-1,1]^N.
\end{eqnarray}

For convenience, we let $\widehat{\varphi}(n,n-1,y) := F_{a,b}(y) - L_{n-1}(y)$. Then, clearly $\varphi(n,n-1,y) = \overline{\sigma} \circ \widehat{\varphi}(n,n-1,y)$.  If $\| \widehat{\varphi}(n,n-1,y) - \widehat{\varphi}(n,n-1,z) \|_{\infty}  \le \delta \| y - z \|_{\infty}$   for all  $y,z, \in [-1,1]^N$ and \eqref{ineq_sigma} holds, then \eqref{eq_mainineq} holds.

We next proceed to prove $\| \widehat{\varphi}(n,n-1,y) - \widehat{\varphi}(n,n-1,z) \|_{\infty}  \le \delta \| y - z \|_{\infty}$ for all  $y,z, \in [-1,1]^N$. 
Recall that $\widehat{\varphi}(n,n-1,y) = (F_{a,b}(y) - L_{n-1}(z))$. We first consider the first term on the right hand side, i.e., $F_{a,b}(y)$ and prove an upper bound for $\| F_{a,b}(y) - F_{a,b}(z) \|_{\infty}$ using the mean value theorem.

Since $f_{a,b}$ is differentiable and has a continuous derivative, by applying the mean value theorem on the domain $[-1,1]$ we infer:
\begin{equation} \label{eq_MVT}
\mid f_{a,b}(y_i) - f_{a,b}(z_i)\mid \: \le \: \sup_{s\in (-1,1)} (\mid f'_{a,b}(s) \mid) \;\cdot  \mid y_i - z_i \mid \mbox{ for all } \: y_i,z_i\in [-1,1]
\end{equation} 
by hypotheses, $a < \frac{\lambda + b}{3}$ and $0< b < \lambda < 1$. If $s\in(-1,1)$, it is straightforward to deduce the following equivalent statements: 
$a < \frac{\lambda + b}{3}$ $\Longleftrightarrow$ $3 a s^2 < \lambda + b$ for all $s\in (-1,1)$ $\Longleftrightarrow$  
$-\lambda + b  < 3 a s^2 < \lambda + b$  for all $s\in (-1,1)$  (since $b < \lambda$)  $\Longleftrightarrow$  $-\lambda <  3 a s^2 - b < \lambda$  for all $s\in (-1,1)$  $\Longleftrightarrow$  $\mid f'_{a,b}(s) \mid < \lambda$ for all $s\in (-1,1)$ since the derivative $f^{'}_{a,b}(s) = 3as^2 - b$.

Since $f^{'}_{a,b}$ is continuous, $\mid f'_{a,b}(s) \mid < \lambda$ for all $s\in (-1,1)$ implies
$\mid f'_{a,b}(s) \mid \le \lambda$ for all $s\in [-1,1]$. 
 Using this in \eqref{eq_MVT}, it follows that 
\begin{equation} \label{eq_MVTineq}
\mid f_{a,b}(y_i) - f_{a,b}(z_i)\mid \: \le \lambda \: \mid y_i - z_i  \mid \: \: \: \mbox{ for all } \: y_i,z_i\in [-1,1]. 
\end{equation}  
Now, 
\begin{eqnarray} \nonumber
\| F_{a,b}(y) - F_{a,b}(z) \|_{\infty} & = & 
\max_{1 \le i \le N} \left(\mid f_{a,b}(y_i) - f_{a,b}(z_i)\mid \right), \\
& \stackrel{\mbox{\eqref{eq_MVTineq}}}{\le} & \max_{1 \le i \le N} \lambda \: \mid y_i - z_i  \mid ,\nonumber \\
& = & \lambda \: \: \| y- z \|_\infty. 
\label{eq_firstineq} 
\end{eqnarray}
Let $\epsilon > 0$ be such that $\delta= \lambda + \epsilon$. Since $L_{n-1}(y) = \overline{\log}(V(n-1)\cdot (d+y))$, the Jacobian $L^{'}_{n-1}$ evaluated at $y$ is 
a diagonal matrix and the $i^{\mbox{th}}$ diagonal element of this matrix is found to be 
\begin{equation} \label{eq_derivative}
\frac{\partial}{\partial y_i}\left( \sum_{j\not= i} \log(V_{ij}(n-1) (d+y_j))\right)
  = \sum_{j\not= i} \frac{1}{(d+y_j)}.
\end{equation}
The following deductions give an upper bound for $\| L_{n-1}(y) - L_{n-1}(z) \|_{\infty}$:
\begin{eqnarray}
\| L_{n-1}(y) - L_{n-1}(z) \|_{\infty} & \le  & \| L_{n-1}(y) - L_{n-1}(z) \|_{2}, \nonumber \\
& \stackrel{\mbox{\eqref{ineq_MVT}}}{\le} & \sup_{s \in int(I^N)} \| L_{n-1}^{'}(s) \|_2 \| y -z\|_2, \nonumber \\
& \le & \sqrt{N} \sup_{s \in int(I^N)} \|  L_{n-1}^{'}(s) \|_\infty \: \: \sqrt{N} \: \| y -z\|_\infty, \nonumber \\
& \stackrel{\mbox{\eqref{eq_derivative}}}{\le} & N \max_{1\le i \le N} \left(\sum_{j\not=i} \frac{1}{d+s_j}\right) \| y -z\|_\infty, \nonumber \\
& \le & \epsilon \: \| y -z\|_\infty \: \: \:\mbox{ (for all } d \mbox{ sufficiently large)}. \label{eq_secondineq}
\end{eqnarray}

Using \eqref{eq_firstineq} and \eqref{eq_secondineq} together,  we get $\| \widehat{\varphi}(n,n-1,y) - \widehat{\varphi}(n,n-1,z) \|_{\infty}  \le (\lambda + \epsilon) \| y - z \|_{\infty} =  \epsilon \| y - z \|_{\infty} $  for all  $y,z, \in [-1,1]^N$. 

Finally, \eqref{ineq_sigma} and \eqref{eq_secondineq} yields the desired bound in \eqref{eq_mainineq}.
Applying \eqref{eq_mainineq} iteratively, we get for any $j \ge 1$, $\| \varphi(n,n-j,y) - \varphi(n,n-j,z) \|_{\infty} \le \delta^j \| y -z\|_\infty $.  Denoting $\delta_j := \delta^j$ in the inequality  in the statement of Lemma~\ref{Lemma_backward_contraction}  it follows that there is exactly one entire-solution for $\varphi$. This proves the theorem.

\end{document}